\documentclass[twoside,11pt]{article}

\usepackage{jmlr_to_arxiv}

\usepackage{hyperref}       
\hypersetup{hidelinks}
\usepackage{url}            
\usepackage{booktabs}       
\usepackage{amsfonts}       
\usepackage{amsmath}
\usepackage{amssymb}
\usepackage{nicefrac}       
\usepackage{microtype}      
\usepackage{xcolor}         
\usepackage{csquotes}

\usepackage{paralist, amsmath, amssymb, graphicx}
\usepackage{amssymb}
\usepackage{setspace}
\usepackage{mathtools}
\usepackage{nccmath}
\usepackage{xcolor}

\usepackage{nicefrac}
\usepackage{stmaryrd}
\usepackage{bm}
\usepackage{cancel}
\usepackage{multirow}

\usepackage{xfrac}
\usepackage{mathrsfs}
\usepackage{amscd}
\usepackage{dsfont}
\usepackage{tikz}
\usepackage{epstopdf}
\usepackage{float}
\usepackage{enumerate}
\usepackage{bm}
\usepackage{chngcntr} 
\usepackage{makecell}
\usepackage{placeins}
\usepackage{algorithm}
\usepackage{algorithmic}
\usepackage{multicol}
\usepackage[colorinlistoftodos]{todonotes}
\usepackage{comment}
\usepackage{enumitem}

\newcommand{\NN}{\mathbb{N}}

\newcommand{\EE}{\mathbb{E}}

\newcommand{\LL}{\mathbb L}

\newcommand{\PP}{\mathbb P}

\newcommand{\la}{\lambda}
\newcommand{\sa}{\sigma}

\renewcommand{\l}{\left}
\renewcommand{\r}{\right}

\newcommand{\norm}[1]{\left\lVert#1\right\rVert}
\newcommand{\normbig}[1]{\big\lVert#1\big\rVert}
\newcommand{\green}{\color{black}}

\DeclareMathOperator{\prox}{prox}
\DeclareMathOperator{\vect}{vec}

\newcommand{\cnorm}[1]{\lVert #1 \rVert_{\LL^2}}
\newcommand{\coef}{\mathcal{B}}

\jmlrheading{1}{x}{1-48}{4/00}{10/00}{xxx}{xxx xxx}

\ShortHeadings{FAStEN}{Boschi et al.}
\firstpageno{1}

\begin{document}


\title{\texttt{FAStEN}: An Efficient Adaptive Method for Feature Selection and Estimation in High-Dimensional Functional Regressions}


\author{\name Tobia Boschi$^*$ \email tobia.boschi@ibm.com \\
       \addr Department of Statistics, Penn State University\\
       \addr IBM Research Europe, Dublin
       \AND
       \name Lorenzo Testa$^*$ \email ltesta@andrew.cmu.edu \\
       \addr Department of Statistics and Data Science, Carnegie Mellon University\\
       \addr L'EMbeDS, Sant'Anna School of Advanced Studies 
       \AND 
       \name Francesca Chiaromonte \email fxc11@psu.edu \\
       \addr Department of Statistics, Penn State University\\
       \addr L'EMbeDS, Sant'Anna School of Advanced Studies
       \AND
       \name Matthew Reimherr \email mlr36@psu.edu \\
       \addr Department of Statistics, Penn State University
       \AND
       \addr $\bm{^*}$ $\text{Equally contributing authors}$
       }

\editor{xxxx}

\maketitle


\begin{abstract}
Functional regression analysis is an established tool for many contemporary scientific applications. Regression problems involving large and complex data sets are ubiquitous, and feature selection is crucial for avoiding overfitting and achieving accurate predictions. We propose a new, flexible and ultra-efficient approach to perform feature selection in a sparse high dimensional function-on-function regression problem, and we show how to extend it to the scalar-on-function framework. 
Our method, called \texttt{FAStEN}, combines functional data, optimization, and machine learning techniques to perform feature selection and parameter estimation simultaneously. We exploit the properties of Functional Principal Components and the sparsity inherent to the Dual Augmented Lagrangian problem to significantly reduce computational cost, and we introduce an adaptive scheme to improve selection accuracy. In addition, we derive asymptotic oracle properties, which guarantee estimation and selection consistency for the proposed \texttt{FAStEN} estimator.
Through an extensive simulation study, we benchmark our approach to the best existing competitors and demonstrate a massive gain in terms of CPU time and selection performance, without sacrificing the quality of the coefficients' estimation. The theoretical derivations and the simulation study provide a strong motivation for our approach. Finally, we present an application to brain fMRI data from the AOMIC PIOP1 study.
Complete \texttt{FAStEN} code is provided at \href{https://github.com/IBM/funGCN}{\texttt{https://github.com/IBM/funGCN}}.
\end{abstract}



%
%
\section{Introduction}
\label{sec:intro}

In contemporary scientific applications, high-dimensional regressions are often handled by imposing sparsity constraints, motivated by the assumption that the explanation of complex phenomena is often reducible to a small number of underlying ``active'' features. In this framework, feature selection methods have a crucial role in pinpointing such features. Among many approaches to feature selection, the most famous and broadly used is the Lasso \citep{tibshirani1996regression}. The Lasso induces sparsity by adding an $l_1$ regularization constraint to the least squares loss function and has proven effective in a variety of scenarios. However, in situations where features exhibit strong linear associations, the Lasso may fail to produce statistically consistent selections \citep{zhao2006model}. 
To deal with such scenarios, extensions have been proposed, such as the Elastic Net \citep{zou2005regularization}, which combines the Lasso $l_1$ regularization with a Ridge $l_2$ regularization. Moreover, in cases where prior knowledge exists on a group structure among the features, such structure can be exploited through a group penalty \citep{yuan2006model}. 

Functional data analysis (FDA) is a less well-known but fast growing field of statistics that deals with structured data in the form of curves and surfaces. The integration between feature selection methods and FDA has received increasing attention in recent years, due to the growing availability of data sets comprising large numbers of functional objects. In fact, \cite{matsui2011variable, gertheiss2013variable, fan2015functional, zhang2022subgroup} study feature selection in scalar-on-function regression settings, where the response is a scalar and the features are functional. \cite{mclean2014functional, chen2016variable, fan2016high, barber2017function, parodi2018simultaneous} study the function-on-scalar case, where the response is functional and the features are scalars.
Notably, less attention has been paid to the more complex case of function-on-function regression \citep{ivanescu2015penalized, qi2018function, wang2020sparse, cai2021variable, centofanti2022smooth}, despite the fact that this type of regression has immediate and important applications in several scientific domains -- from ecology \citep{luo2021restricted}, to epidemiology \citep{boschi2021functional, boschi2023contrasting}, to the neurosciences \citep{qi2018function}. The function-on-function feature selection problem is computationally intricate, as it deals with data whose nature is over a continuum and does not assume any lag structure on the relationship between features and response (thus, each regression coefficient is a surface). However, by exploiting recent advances in optimization, we extend state-of-the-art methodology for feature selection in function-on-scalar regressions \citep{boschi2021highly} to tackle the function-on-function case, even in high- and ultra high-dimensional settings. 
The function-on-function regression model with $n$ observations and $p$ features takes the form \citep{ramsay2005, kokoszka2017introduction}:
\begin{align}
\small
\begin{split}
\label{eq:fof_problem}
    \mathcal{Y}_i(t) = \sum_{j=1}^p\int_\mathcal{S}\mathcal{B}_j(t,s)\mathcal{X}_{ij}(s)ds + \epsilon_i(t) \qquad i = 1, \dots, n
\end{split}
\end{align}
where the $\mathcal{Y}_i$ are centered functional responses (with $0$ mean function), the $\mathcal{X}_{ij}$ are standardized functional features (with a mean of 0 and a standard deviation of 1 across the entire domain), the $\mathcal{B}_j$ are coefficient surfaces to be estimated, and the $\epsilon_i$ are i.i.d.~errors, independent of the $\mathcal{X}$'s, with $0$ mean function and a common variance operator.
Let $\mathcal T = [a,b]$ and $\mathcal S = [a',b']$ be two closed bounded intervals. We assume that $\mathcal{Y}_i \in \mathbb{L}^2(\mathcal T)$, $\mathcal X_{ij} \in \mathbb{L}^2(\mathcal S)$, and $\mathcal{B}_j \in \mathbb{L}^2(\mathcal T) \times \mathbb{L}^2(\mathcal S)$. 
We use the notation $\mathcal{Y} = \left[\mathcal{Y}_1 ~ | ~ \dots ~ | ~ \mathcal{Y}_n \right]^T$ to denote the collection of response curves, $\coef = \left[\mathcal{B}_1 ~ | ~ \dots ~ | ~ \mathcal{B}_p \right]^T$ to represent the set of $p$ coefficient surfaces, and $\mathcal{X}_j = \left[\mathcal{X}_{1j} ~ | ~ \dots ~ | ~ \mathcal{X}_{nj} \right]^T$ to indicate the set of $n$ curves of the $j$-th feature. 

In this paper, we propose a novel method to perform feature selection for 
(\ref{eq:fof_problem}), which we call Functional Adaptive feature SelecTion with Elastic Net penalty (\texttt{FAStEN}).
For a generic vector of functions $f = \left[f_1 ~ | ~ \dots ~ | ~ f_n \right]^T$  and surfaces $g = \left[g_1 ~ | ~ \dots ~ | ~ g_n \right]^T$, with $f_i \in \mathbb L^2(\mathcal{T})$, $g_i \in \mathbb L^2(\mathcal{T}) \times \mathbb L^2(\mathcal{S})$ $\forall i$, we define the $\mathbb{L}^2$-norms as $\lVert f\rVert_{\mathbb{L}^2} = ( \sum_{i=1}^n \lVert f_i\rVert_{\mathbb{L}^2}^2 )^{1/2}$ and  $\lVert g\rVert_{\mathbb{L}^2 \times \mathbb{L}^2} = ( \sum_{i=1}^n \lVert g_i\rVert_{\mathbb{L}^2 \times \mathbb{L}^2}^2 )^{1/2}$, respectively. 
In particular, $\lVert f_i\rVert_{\mathbb{L}^2}^2= \langle f_i, f_i\rangle_{\mathbb{L}^2} = \int_\mathcal{T} f_i^2$ and 
$\lVert g_i\rVert_{\mathbb{L}^2 \times \mathbb{L}^2}^2= \langle g_i, g_i\rangle_{\mathbb{L}^2 \times \mathbb{L}^2} = \int_\mathcal{S}\int_\mathcal{T} g_i^2$. Then, the \texttt{FAStEN} optimization problem is given by
%
\begin{align}
\small
\begin{split}
    \label{eq:functional_minimization_problem}
    \min_{\mathcal{B}_1, \dots, \mathcal{B}_p} ~ \Bigg[ 
    \frac{1}{2} \Big\lVert~\mathcal{Y}-\sum_{j=1}^p\int_\mathcal{S}\mathcal{B}_j(t,s)\mathcal{X}_j(s)ds~\Big\rVert_{\mathbb{L}^2}^2 +
    \sum_{j=1}^{p} \omega_j \left(
     \lambda_1 \lVert \mathcal{B}_j\rVert_{\mathbb{L}^2 \times \mathbb{L}^2}+ 
     \frac{\lambda_2}{2}\lVert \mathcal{B}_j \rVert_{\mathbb{L}^2 \times \mathbb{L}^2}^2 
    \right)  \Bigg]
\end{split}
\end{align}
with two different penalties added to the least squares loss. The parameters $\lambda_1$ and $\lambda_2$ control the importance of the penalties with respect to the least squares loss, and the weights $\omega_j$ extend the \textit{adaptive} Lasso \citep{zou2006adaptive} and Elastic Net \citep{zou2009adaptive} to function-on-function regression, improving selection and estimation of coefficient surfaces.

To solve \eqref{eq:functional_minimization_problem}, first we approximate the functional variables by a matrix representation, then we develop a new, very efficient method to carry out the minimization. 
The matrix representation is achieved through the \textit{Functional Principal Components} (FPC)  \citep[see e.g.][]{horvath2012inference}; response and features curves are replaced by FPC scores. The minimization is performed through a \emph{Dual Augmented Lagrangian} (DAL) algorithm 
which exploits the sparsity of the problem and the sparsity induced by the \emph{Hessian matrix} information to greatly reduce computational cost. DAL was first introduced by \cite{tomioka2009dual, li2018} to solve the Lasso problem and by \cite{tomioka2011super, boschi2020efficient} to solve the Elastic Net problem. \cite{boschi2021highly} presented a version of DAL to solve the Group Elastic Net problem and perform feature selection in function-on-scalar regressions. In this paper, we introduce a new adaptive version of DAL to perform feature selection in function-on-function regressions -- which requires that we carefully modify and expand a set of mathematical operators to deal with a larger and more complex optimization. 
In the Appendix, we also provide details on how to implement DAL in the \emph{scalar-on-function} framework \citep[see e.g.][]{kokoszka2017introduction}. This work, together with \cite{boschi2021highly}, outlines a novel and highly efficient feature selection methodology applicable to all the most common functional regression settings.   

We implement an efficient version of \texttt{FAStEN}, written in \texttt{python}, and benchmark it against \texttt{FRegSigCom} \citep{qi2018function}, written in \texttt{R} with backend in \texttt{C++}. 
Since function-on-function feature selection has not yet been intensively studied, this was the only existing method we found to deal with high-dimensional scenarios ($p \gg n$) and with publicly available code.
%
%
Our simulations demonstrate that \texttt{FAStEN} is at least three orders of magnitude faster and more accurate in terms of false positives than \texttt{FRegSigCom}, while possessing similar accuracy in terms of coefficient estimation and out-of-sample prediction error. 
We also apply \texttt{FAStEN} to data from the Amsterdam Open MRI Collection (AOMIC) PIOP1 study \citep{snoek2021amsterdam}, which investigates activation of brain voxels in response to external stimuli. Specifically, we examine the association between the activation functions of tens of thousands of brain voxels and heart rate variability functions recorded on the study subjects while dealing with the external stimuli of the so-called anticipation task.

The remainder of the paper is organized as follows. In Section \ref{sec:method}, we describe how to approximate the functional optimization problem via FPCs and how to solve it with our DAL-based approach. In Section \ref{sec:theoretical_results}, we provide theoretical guarantees on the asymptotic oracle properties of the \texttt{FAStEN} estimator. 
In Section \ref{sec:50shadows}, we detail various specifics on the algorithmic implementation. In Section \ref{sec:sim_and_application}, we investigate \texttt{FAStEN} performances through an extensive simulation study. 
In Section \ref{sec:application}, we apply \texttt{FAStEN} to the AOMIC PIOP1 data. Finally, in Section \ref{sec:conclusion}, we provide final remarks and discuss future developments. 
Proofs and extensions of theoretical results, as well as results for additional simulation scenarios, can be found in the Appendix. 
Complete \texttt{FAStEN} code is available on \texttt{GitHub} at: \href{https://github.com/IBM/funGCN} {\texttt{https://github.com/IBM/funGCN}}.

%
%

\section{Methodology}
\label{sec:method}

\subsection{Functional Principal Components' matrix representation}
\label{subsec:coeff_representation}

One of the most common methods to work with model \eqref{eq:fof_problem}
is to use a matrix representation obtained expressing functional variables as linear combinations of basis functions \citep{ramsay2005, horvath2012inference, ivanescu2015penalized, beyaztas2020function}. 
The design of \texttt{FAStEN} allows one to work with different basis for the $\mathcal X's$ and the $\mathcal Y$. In the following, as in standard FPC regression \citep{reiss2007functional}, we represent each functional variable using its own FPCs.
We now show how one can use the FPCs in order to obtain a matrix representation of the objective function in \eqref{eq:functional_minimization_problem}.

First, let us consider the simple case with just one feature, $\mathcal X_1$. Let $\Upsilon = \left[\Upsilon_1 ~ | ~ \dots ~ | ~ \Upsilon_K \right]$ be the matrix containing the first $K$ FPCs of $\mathcal Y$ estimated from the sample, and $Y \in \mathbb{R}^{n \times K}$ be the matrix of the scores of $\mathcal Y$ with respect to $\Upsilon$, i.e. $Y_{ik} = \langle \mathcal Y_i, \Upsilon_k \rangle_{\mathbb{L}^2}$. Similarly, let $\Xi_1 = \left[\Xi_{11} ~ | ~ \dots ~ | ~ \Xi_{1L} \right]$ be the matrix containing the first $L$ FPCs of $\mathcal{X}_1$ estimated from the sample, and $X_{[1]} \in \mathbb{R}^{n \times L}$ be the matrix of the scores of $\mathcal{X}_1$ with respect to $\Xi_1$, i.e. $X_{[1](i\ell)} = \langle \mathcal{X}_{i1}, \Xi_{1\ell} \rangle_{\mathbb{L}^2}$.
Let $B_{[1]} \in \mathbb{R}^{K \times L}$ be the score matrix of $\mathcal B_1$ with respect to $\Upsilon\Xi_1^T$;
that is, $B_{[1](k\ell)} = \langle \mathcal B_1, \Upsilon_{k} \otimes  \Xi_{1\ell} \rangle_{\mathbb{L}^2 \times \mathbb{L}^2} = \int\int \mathcal B_1(t, s) \Upsilon_{k}(t)\Xi_{1\ell}(s)\,dt\,ds$. Then, we can approximate $\mathcal Y(t) \approx Y\Upsilon(t)^T, ~ \mathcal X_1(s) \approx X_{[1]}\Xi_1(s)^T, ~ \mathcal B_1(t,s) \approx \Upsilon(t)B_{[1]}\Xi_1(s)^T$ and rewrite model~\eqref{eq:fof_problem} as
 \begin{align*}
 \small
 \begin{split}
     Y\Upsilon(t)^T \approx X_{[1]} \left(\int_{\mathcal{S}} \Xi_1(s)^T\Xi_1(s)~ds\right) B_{[1]}\Upsilon(t)^T \ .
 \end{split}
 \end{align*}
Since for an orthonormal basis $\int \Xi_1(s)^T\Xi_1(s)\,ds = I_L$, the identity matrix of order $L$, we can multiply both sides of the equation by $\Upsilon$ and integrate over $t$. Again, $\int \Upsilon(t)^T\Upsilon(t)dt = I_K$. 
This leads us to $Y \approx X_{[1]} B_{[1]}$. The extension to $p$ predictors is straightforward.
 
Let us indicate the design matrix as $X = \left[X_{[1]} | \dots | X_{[p]} \right] \in \mathbb{R}^{n \times pL}$, the coefficient matrix as $B = \left[B_{[1]} | \dots | B_{[p]} \right]^T \in \mathbb{R}^{pK\times L}$, and the matrices restricted to the scores of the $j$-th feature and the $j$-th coefficient surface as $X_{[j]}\in \mathbb{R}^{n \times L}$ and $B_{[j]}\in \mathbb{R}^{K \times L}$, respectively. The function-on-function linear model in \eqref{eq:fof_problem} can be approximated as $ Y \approx X B $. To provide a matrix representation of the objective function in~\eqref{eq:functional_minimization_problem}, we need to relate the ${\mathbb{L}^2}$ functional norm and the standard $l_2$ matrix norm, i.e.~the Frobenius norm -- which we indicate with $\lVert \cdot \rVert_2$. Because $\Upsilon$ and $\Xi_j$, $j=1,\dots,p$, are orthonormal, for a generic function $f \in \mathbb L^2([0,1])$ and a generic surface $g \in \mathbb L^2([0,1])\times \mathbb L^2([0,1])$ one has
 \begin{align}
 \small
 \begin{split}
 \label{eq:fpca_prop}
     &\lVert f \rVert_{\mathbb{L}^2}^2 = 
     \sum_{k=1}^\infty \langle f, \Upsilon_k\rangle_{\mathbb{L}^2} \ , \ \ 
     \lVert g \rVert_{\mathbb{L}^2 \times \mathbb{L}^2}^2 = \sum_{k=1}^\infty \sum_{\ell=1}^\infty  \big\langle \langle g, \Xi_{j\ell}\rangle_{\mathbb{L}^2}, \Upsilon_{k} \big\rangle_{\mathbb{L}^2} \ .
 \end{split}
 \end{align}
 These equations allow us to approximate the ${\mathbb{L}^2}$ functional norm with the $l_2$ Frobenius norm for both vectors and matrices. 
 We are now ready to express \eqref{eq:functional_minimization_problem} in matrix form as
 \begin{equation}
 \small
     \label{eq:matrix_minimization_problem}
     \min_{B} \frac{1}{2}\lVert Y-XB\rVert_{2}^2 + \sum_{j=1}^{p} \omega_j \left( \lambda_1 \lVert B_{[j]}\rVert_{2}+\frac{\lambda_2}{2} \lVert B_{[j]}\rVert_{2}^2 \right) \ .
 \end{equation}
Like in an Elastic Net \citep{zou2005regularization}, the optimization combines two penalties; the first is non-differentiable and creates sparsity, and the second (Ridge-like) is differentiable and controls variance inflation due to multicollinearity. The introduction of a Ridge-like term leads to differentiable operators which speed up the convergence of the optimization algorithm. Once an estimate $\check B$ of the score matrix is obtained from~\eqref{eq:matrix_minimization_problem}, one can recover the coefficient surfaces as
\begin{equation*}
\small
\label{eq:from_B_to_funB}
\hat{\mathcal B}_j(t,s) = \Upsilon(t)\check B_{[j]}\Xi_j(s)^T \ \ \ j=1,\dots,p \ .
\end{equation*}

In general, FPCs are a means to control, and appropriately calibrate, the dimension of the problem. On the one hand, low $K$ and $L$ may make the approximation too coarse, leading to the loss of important signals and reducing the quality of results. On the other hand, high $K$ and $L$ may retain too much noise, leading to less accurate estimated surfaces. 
FPCs are also a means to compute integrals more efficiently, as they can be decomposed in sums over the squares of the FPC scores;
see~\eqref{eq:fpca_prop}.

\subsection{Dual formulation}
\label{subsec:dual}

Here, we introduce the \emph{dual formulation} of the objective function in~\eqref{eq:matrix_minimization_problem}.  The core idea behind the DAL is to minimize the \emph{Augmented Lagrangian} function associated to the \emph{dual} problem in order to find its optimal solution.
First, notice that \eqref{eq:matrix_minimization_problem} can be written as
%
\begin{equation}
\small
    \label{eq:primal}
    \tag{P}
    \min_B h(XB) + \pi(B)
\end{equation}
where 
$h(XB)= \frac{1}{2}\lVert Y-XB\rVert_{2}^2$ 
is the least squares loss function and 
$\pi(B) = \sum_{j=1}^{p} \omega_j 
( \lambda_1 \lVert B_{[j]}\rVert_{2}$
$+\frac{\lambda_2}{2} \lVert B_{[j]}\rVert_{2}^2 )
$ 
is the adaptive Elastic Net-type penalty. A possible dual formulation of the primal 
\eqref{eq:primal} is given by \citep[Sec.~5.1.6]{boyd2004convex}
%
\begin{equation}
\small
    \label{eq:dual}
    \tag{D}
    \min_{V,Z} h^*(V) + \pi^*(Z)\quad s.t.\quad X^TV+Z=0 \ \ 
\end{equation}
%
where $V\in\mathbb{R}^{n\times K}$ and $Z\in\mathbb{R}^{pK\times L}$ are the matrix dual variables. 
Using the same notation 
introduced for $B$, we can write $Z = \left[Z_{[1]} | \dots | Z_{[p]} \right]^T$, where 
$Z_{[j]}\in \mathbb{R}^{K \times L}$ is the sub-matrix of $Z$ associated with the $j$-th feature. 
$h^*$ and $\pi^*$ are the Fenchel-conjugate functions \citep{fenchel1949conjugate} of $h$ and $p$, respectively. The form of $h^*$ is known in the literature: $h^*(V) = \frac{1}{2}\lVert V\rVert_2^2 + \langle Y, V\rangle$ \citep{dunner2016primal}, where $\langle \cdot, \cdot \rangle$ is the Frobenius inner product for matrices. The form of $p^*$ is given in the next Proposition, which extends 
results in \cite{boschi2021highly} from the case of vectors to the case of matrices (see Appendix Section \ref{subsec:proof_prop1} for a proof).
\begin{proposition}
\label{prop:p_star}
    Given the problem~\eqref{eq:dual}, the function $\pi^*$ has the form 
    \begin{align}
    \small
    \begin{split}
    	\label{eq:p_star}
        	\pi^*(Z) = \sum_{j=1}^p \pi^*(Z_{[j]}) = \sum_{j=1}^p (2\omega_j\la_2)^{-1} \left(\big[\norm{Z_{[j]}}_2 - \omega_j\la_1\big]_{+}\right)^2
    \end{split}
    \end{align}
    where $[~\cdot~]_{+}$ is the positive part operator; $[s]_{+} = s$ if $s>0$ and $0$ otherwise. 
\end{proposition}
Notably,  $\pi^* : \mathbb{R}^{pK \times L} \rightarrow \mathbb{R}$ is a differentiable and separable function despite the fact that the objective function in~\eqref{eq:matrix_minimization_problem} is not separable. As we will see in Section~\ref{subsec:dal},
these two properties 
are crucial for preserving the sparsity structure of the DAL-based optimization.

Primal and dual optimality 
can be investigated 
through the \textit{Karush-Kuhn-Tucker} (KKT) conditions associated with the dual problem~\eqref{eq:dual}; namely
%
\begin{equation}
\small
    \label{eq:kkt}
    \nabla h^*(V) - XB = 0 
    \ \ , \ \ 
    \nabla \pi^*(Z) - B = 0 
    \ \ , \ \ 
    X^T V + Z = 0
    \ \ .
\end{equation}
It can be shown that the tuple $(V^{opt}, Z^{opt}, B^{opt})$ solves 
these conditions
if and only if $(V^{opt}, Z^{opt})$ is the optimal solution of \eqref{eq:dual}
and $B^{opt}$ is the optimal solution of \eqref{eq:primal}
\citep[Sec.~5.5.3]{boyd2004convex}. 


\begin{algorithm}[t!]
\caption{\textbf{DAL Method}}
\label{alg:dal}
\small
\begin{algorithmic}

\vspace{0.2cm}
\STATE \textbf{GOAL:} minimize  $\mathcal{L}_\sigma (V,Z,B)$. Start from the initial values $V^0, Z^0, B^0, \sa^0$ \\

\vspace{0.2cm}
\STATE \texttt{WHILE NOT CONVERGED:}
\begin{itemize}
    
    \item[\textbf{(1)}] Given $B^s$, find $V^{s+1}$ and $Z^{s+1}$ which approximately solve:
    \begin{equation}
        \label{eq:inner_subproblem}
         	 \big(V^{s+1}, Z^{s+1}\big) \approx \arg \min_{V,Z} \mathcal{L}_\sa \big(V, Z ~|~ B^s\big)
    \end{equation}
    \vspace{-0.5cm}
    
    \fbox{\begin{minipage}{35em}
    
        \vspace{0.2cm}
        ~ ~ \textbf{Inner sub-problem:} to find $(V^{s+1}$, $Z^{s+1})$ update V, Z \emph{independently}: \\
        
        \vspace{-0.1cm}
        ~ ~ \texttt{WHILE NOT CONVERGED}   
        
        \begin{itemize}
            
            \vspace{0.1cm}
            \item[]  $V^{m+1} = \arg\min_V \mathcal{L}_\sigma (V\,|\,Z^m,B^s)~ \longrightarrow~$ 
            {\footnotesize \textbf{Newton method} (Theorem \ref{th:v_update})}
            
            \vspace{0.1cm}
            \item[] $Z^{m+1} = \arg\min_Z \mathcal{L}_\sigma (Z,|\,V^{m+1},B^s)~ \longrightarrow~$ 
            {\footnotesize \textbf{closed form} (Proposition \ref{prop:z_update})}
            \vspace{0.2cm}
        
        \end{itemize}
    
    \end{minipage}}
    
    \vspace{0.1cm}
    \item[\textbf{(2)}] Update the Lagrangian multiplier $B$ and the parameter $\sa$:
    
    \vspace*{-0.1cm}
    \begin{align}
    \label{eq:update_x}
    \begin{split}
                        &B^{s+1} = B^s - \sa^s \big(X^T V^{s+1} + Z^{s+1}\big) \\
                        &\sa^{s+1} \uparrow \sa^{\infty} \le \infty
    \end{split}
    \end{align}
\end{itemize}
\vspace*{-0.4cm}
\end{algorithmic}
\end{algorithm}
\subsection{DAL algorithm}
\label{subsec:dal}

In this section, we present the \emph{Dual Augmented Lagrangian} (DAL) algorithm, which efficiently solves the optimization problem in \eqref{eq:matrix_minimization_problem}. 
Specifically, we describe here the DAL implementation for function-on-function regression, while
Appendix Section \ref{sec:supp_scalar}
outlines the scalar-on-function regression case
\citep{reiss2017methods}.


We note that the proposed extension addresses an optimization problem distinct from those previously studied in the literature. An easier version of the problem in~\eqref{eq:matrix_minimization_problem} -- without the Ridge penalty and adaptive weights -- was solved by \cite{chen2012sparse} using a subgradient method. This optimization strategy has been shown to be significantly less efficient compared to the DAL method, both in classical \citep{tomioka2011super} and functional \citep{boschi2021highly} settings.
Compared to other DAL implementations, \eqref{eq:matrix_minimization_problem} incorporates functional variables both as response and as predictors.
This increases the dimension and the computational complexity of the problem with respect to the function-on-scalar case \citep{boschi2021highly}.

We need to carefully redefine all the mathematical operators in larger dimensions and, at the same time, preserve the DAL sparsity structure and computational efficiency. 
Moreover, by incorporating weights (the $\omega$'s) in the objective function in \eqref{eq:matrix_minimization_problem}, we create a novel, \emph{adaptive} version of DAL which improves both selection and estimation performance \citep{zou2006adaptive, zou2009adaptive, parodi2018simultaneous}, as we will see in 
Section \ref{sec:sim_and_application}.

Our implementation, summarized in \textbf{Algorithm \ref{alg:dal}}, minimizes the \textit{Dual Augmented Lagrangian} function
\begin{align}
\small
\begin{split}
    \label{eq:augmented_lagrangian}
    \mathcal{L}_\sigma (V,Z,B) =  h^*(V) + \pi^*(Z) - 
     \sum_{j=1}^p \langle B_{[j]}, V^TX_{[j]} + Z_{[j]} \rangle + 
    \frac{\sigma}{2} \sum_{j=1}^p \lVert V^TX_{[j]} + Z_{[j]}\rVert_2^2
\end{split}
\end{align}
where $\sigma>0$. $B$ is both the primal variable in 
\eqref{eq:primal} and the Lagrangian multiplier in
\eqref{eq:augmented_lagrangian}, weighting the importance of a violation of the constraint. 

The core of the algorithm is the routine to solve the inner sub-problem \eqref{eq:inner_subproblem}, while the updates of $B$ and $\sigma$ follow standard optimization rules \citep{li2018}. \cite{tomioka2009dual} show that an approximate solution $(\Bar{V},\Bar{Z})$ of \eqref{eq:inner_subproblem} can be found independently for V and Z, as
%
\begin{equation*}
\small
    \Bar{V} = \arg\min_V \mathcal{L}_\sigma (V\,|\,\Bar{Z},B)
    \ \ , \ \ 
    \Bar{Z} = \arg\min_Z \mathcal{L}_\sigma (Z\,|\,\Bar{V},B)
\end{equation*}
where, with a slight abuse of notation, we indicate by $\mathcal{L}_\sigma(V\,|\,Z,B)$ and $\mathcal{L}_\sigma(Z\,|\,V,B)$ the functions obtained from $\mathcal{L}_\sigma(V,Z,B)$ fixing the parameters $Z$ and $B$, or $V$ and $B$, respectively. 

\paragraph{Z update.}
The following proposition extends the findings of \cite{boschi2021highly} on the \emph{proximal operator} \citep{rockafellar1976, rockafellar1976augmented} of $\pi$ and 
provides a closed form for $\Bar{Z}$ (see Appendix Section \ref{subsec:proof_prop2} for a proof).
\begin{proposition}
\label{prop:z_update}
        %
        %
    
The proximal operator of the penalty function $\pi(B)$ defined in \eqref{eq:primal} is given by 
$\prox_{\sigma \pi}(B) = \big(\prox_{\sigma \pi}(B_{[1]}), \dots, \prox_{\sigma \pi}(B_{[p]}) \big)^T$, 
where for each $j$  
        \begin{equation}
        \small
        	\label{eq:prox_op}
		    \prox_{\sigma \pi}(B_{[j]}) =  (1 + \sa \omega_j \la_2)^{-1} \left[1- \norm{B_{[j]}}_2^{-1} \sa \omega_j \la_1  \right]_{+} B_{[j]}.
	    \end{equation}
Thus, we have
        $ \Bar{Z} = \prox_{\pi^*/\sa} \left(B/\sa - X^T\Bar{V} \right) = B/\sa - X^T\Bar{V} - \prox_{\sa p}\left(B - \sa X^T\Bar{V}\right)/\sa$.
\end{proposition}
Note that $\prox_{\sigma \pi}(B) : \mathbb{R}^{pK \times L} \rightarrow\mathbb{R}^{pK \times L}$. The input argument has a larger dimension than the one in the function-on-scalar case, which deals with ${p \times K}$ matrices.

\paragraph{V update.}
We do not have a closed form for $\bar V$. 
We update $V$ using the Newton Method \citep{nocedal1999numerical}. 
If $\psi(V) = \mathcal{L}_\sigma (V\,|\,\Bar{Z},B)$ indicates the function we want to minimize with respect to $V$, the Newton method update has the form 
%
    $V^{m+1} = V^{m} + uD$,  
%
where $u$ represents the step-size and $D \in \mathbb{R}^{n \times K}$ is the descent direction computed by solving the linear system
\begin{equation}
\small
\label{eq:newton_direction}
    H_{\psi}(V) \vect(D) = -\vect\big(\nabla \psi (V)\big) \ .
\end{equation}
Here, $H_{\psi} \in \mathbb R^{nK \times nK}$ is the Hessian matrix, 
$\nabla \psi \in \mathbb R^{n \times K}$ is the gradient matrix, and 
$\vect \big(\nabla \psi \big) \in \mathbb R^{nK \times 1}$ is the gradient vector obtained stacking all the columns of $\nabla \psi$ (similarly for $D$ and $\vect(D)$).
{\green 
To determine the step size, we implement the \emph{line-search} strategy proposed by \cite{li2018}. Starting with $u=1$, we iteratively reduce it until the following condition is satisfied:
\begin{equation}
\label{eq:linesearch}
	\psi (V^{m+1}) \le \psi (V^m) + \mu u \vect \big(\nabla \psi(V^m) \big)^T \vect (D)
\end{equation}
with $\mu \in (0, 0.5)$. }
The largest computational burden in DAL comes from solving the linear system in \eqref{eq:newton_direction}. 
The following theorem allows us to reduce such burden by greatly reducing the dimension of the system (see Appendix Section \ref{subsec:proof_theorem1} for a proof).
\begin{theorem}
\label{th:v_update}
Let $T=B-\sa X^T V$, $T_{[j]}=B_{[j]} - \sa V^T X_{[j]}$, $\mathcal{J} = \big\{ j~:~ \lVert T_{[j]} \rVert_2  \ge \sa \omega_j \la_1 \big\}$, and $r = |\mathcal{J}|$ be the cardinality of $\mathcal{J}$.
Next, let $ X_\mathcal{J} \in \mathbb{R}^{n \times rL}$ be the sub-matrix of $X$ restricted to the blocks $X_{[j]}$, 
$j \in \mathcal{J}$, and $\hat X_\mathcal{J} = X_\mathcal{J} \otimes_{Kron} I_K$ be the $nK \times rKL$ Kronecker product between $X_\mathcal{J}$ and the $K \times K$ identity matrix.
Next, consider the $KL \times KL$ outer product $\hat T_{[j]} = \vect\left(T_{[j]}\right) \otimes_{outer} \vect\left(T_{[j]}\right)$ and define the squared $KL \times KL$ matrix 
    \begin{equation*}
    \small
    \label{eq:P_j}
        P_{[j]} = (1 + \sa \omega_j \la_2)^{-1} \left( \left(1 -
        \frac{\sa \omega_j \la_1}{ \norm{T_{[j]}}_2} \right) I_{KL} +
        \frac{\sa \omega_j \la_1}{ \norm{T_{[j]}}_2^3} 
        \hat T_{[j]} \right) \ .
    \end{equation*}
Finally, let $Q_\mathcal{J} \in \mathbb{R}^{rKL \times rKL}$ be the block-diagonal matrix formed by the blocks $P_{[j]}$, $j \in \mathcal{J}$.
Then
  \begin{align*}
    \small
    \label{eq:psi_y_z_bar}
    \begin{split}
        &(i)~ \  \ \psi(V) = h^*(V) + \frac{1}{2\sa}  \sum_{j=1}^p 
        \left( \left( 1 + \sa \omega_j \la_2 \right) \normbig{\prox_{\sa \pi}\big(T_{[j]}\big) }_2^2 - \norm{B_{[j]}}_2^2 \right)\,; \\ 
        &(ii)~ \ \nabla \psi(V) = V + Y - X \prox_{\sa \pi}(T)\,; \\
        &(iii)~H_{\psi}(V) = I_{nK} + \sa \hat X_\mathcal{J} Q_\mathcal{J} \hat X_\mathcal{J}^T \ .
    \end{split}
    \end{align*}
\end{theorem}
%

\paragraph{Computational cost.}
To reduce the dimension of the problem, and thus the computational burden, our theorem exploits the sparsity structure inherent to the second order information in the Augmented Lagrangian function. At each iteration, DAL selects a subset of active features $\mathcal{J}$ with cardinality $r$, reducing the total cost of solving the linear system (matrix multiplication and \emph{Cholesky} factorization) from $\mathcal{O} \l(nK^3(n^2 + p^2L^2 + npL) \r)$ to $\mathcal{O} \l( nK^3(n^2 + r^2L^2 + nrL) \r)$.
The number of active features $r$ decreases along the iterations of the algorithm, and in sparse scenarios $r$ is usually much smaller than $n$.
Therefore, the cost of the system can be further reduced performing the inversion through the \emph{Sherman-Morrison-Woodbury} formula \citep{van1983matrix},
whereby {\small $\left(I_{nK} + \sa \hat X_\mathcal{J} Q_\mathcal{J} \hat X_\mathcal{J}^T \right)^{-1}$} is equivalent to
%
{\small
	   $I_{nK} -  \hat  X_\mathcal{J}  \left( \l( \sa Q_\mathcal{J} \r)^{-1} + \hat X_\mathcal{J}^T \hat X_\mathcal{J} \right)^{-1} \hat X_\mathcal{J}^T$},
%
which allows one to factorize an $rK \times rK$ matrix. The total cost is now $\mathcal{O} \l( rK^3L(L^2 + L^2r^2 + n^2 + rnL) \r)$.
Notably, this cost does not depend on the total number of features $p$, but just on the number of active features $r$. 
$p$ can increase without affecting the efficiency of the algorithm, as long as sparsity holds, i.e.~$r$ is small. 
The left panel of Figure \ref{fig:complexity_adaptive} illustrates this property;
the CPU time does not increase with $p$, it increases with $r$.
Finally, when comparing the computational cost between the function-on-function case presented here and the function-on-scalar one presented in \cite{boschi2021highly}, the main difference is that solving the linear system depends on both $K^3$ and $L^3$, rather than just $K^3$. 
This makes small values of $K$ and $L$ -- i.e.,  representing both the response and the features using a few FPCs -- even more crucial for improving computational efficiency.

\paragraph{Stopping criteria.} 
%
DAL convergence has been already explored in previous literature \citep{tomioka2009dual, li2018, boschi2020efficient}. 
The convergence analysis for DAL with an 
Elastic Net penalty relies on the differentiability of $\pi^*$.
{\green As showed and detailed in Appendix Section \ref{sec:supp_conv}, both the DAL algorithm and the routine for the inner sub-problem have a super-linear convergence rate;}
\texttt{FAStEN} converges in very few iterations. 
Following previous literature, to determine the convergence of the algorithm and of the inner sub-problem routine (i.e. the iterative update of $Z$ and $V$), we check the standardized residuals of the third and first KKT equations in \eqref{eq:kkt}, respectively. 
Using $i$ 
to index the rows of a matrix,
these are
\begin{align*}
\small
\begin{split}
    \text{res}(\text{kkt}_3) = \frac{ \sum_{j=1}^p\norm{V^T X_{[j]} + Z_{[j]}}_2} {1 +  \sum_{i=1}^n \norm{V_{i}}_2 +  \sum_{j=1}^p \norm{Z_{[j]}}_2}
    \ \ \ , \ \ \
    \text{res}(\text{kkt}_1) = \frac{ \sum_{i=1}^n\norm{(V + Y - XB)_{i}}_2} {1 +  \sum_{i=1}^n \norm{Y_{i}}_2 + \sum_{j=1}^p \norm{X_{[j]}}_2} \ \ .
\end{split}
\end{align*}
%

%
%
{\green 
\section{Oracle properties}
\label{sec:theoretical_results}

In this section, we establish
asymptotic oracle properties 
for the \texttt{FAStEN} estimator. 
These properties refer to the ability of \texttt{FAStEN} to consistently identify the correct set of active features and estimate the coefficients as accurately \textit{as if} the true underlying model were known in advance.
The results we provide, while relying on specific regularity conditions, hold in a far more general setting than the one 
used in prior Sections, and require us to introduce additional notation. 
First, according to Mercer's theorem, we can represent the covariance functions of $\mathcal{Y}(t)$ and $\mathcal{X}_j(s)$, $j=1,\dots, p$, respectively, as $C_{\mathcal{Y}}(t_1, t_2) = \sum_{k=1}^\infty \mathring{\upsilon}_{k} \mathring{\Upsilon}_k(t_1) \mathring{\Upsilon}_k(t_2)$ and $C_{\mathcal{X}_{j}}(s_1, s_2) = \sum_{\ell=1}^\infty \mathring{\xi}_{j\ell} \mathring{\Xi}_{j\ell}(s_1) \mathring{\Xi}_{j\ell}(s_2)$, where $\{\mathring{\upsilon}_{k}:\,k\in\NN\}$ and $\{\mathring{\xi}_{j\ell}:\,\ell\in\NN\}$ are eigenvalue sequences associated to the orthonormal eigenfunctions $\{\mathring{\Upsilon}_k:\,k\in\NN\}$ and $\{\mathring{\Xi}_{j\ell}:\,\ell\in\NN\}$ spanning $\LL^2(\mathcal{T})$ and $\LL^2(\mathcal{S})$. 
For sample curves, according to the Karhunen-Loève expansion, we have $\mathcal{Y}_i(t) = \sum_{k=1}^\infty \mathring{Y}_{ik} \mathring{\Upsilon}_k(t)$ and $\mathcal{X}_{ij}(s) = \sum_{\ell=1}^\infty \mathring{X}_{[j](i\ell)} \mathring{\Xi}_{j\ell}(s)$, where $\mathring{Y}_{ik}$ and $\mathring{X}_{[j](i\ell)}$ are uncorrelated random variables with zero mean and variances $\EE[\mathring{Y}_{ik}^2] = \mathring{\upsilon}_k$ and $\EE[\mathring{X}_{[j](i\ell)}^2] = \mathring{\xi}_{j\ell}$, respectively. Similarly, we can write $\mathcal{B}_j(t,s) = \sum_{k=1}^\infty \sum_{\ell=1}^\infty \mathring{B}_{[j](k\ell)} \mathring{\Upsilon}_k(t) \mathring{\Xi}_{j\ell}(s)$, $j=1,\dots,p$. 
Substituting these expansions in (\ref{eq:fof_problem}), we 
obtain
\begin{equation*}
    \small
    \sum_{k=1}^\infty \mathring{Y}_{ik} \mathring{\Upsilon}_k(t) = \sum_{j=1}^p \sum_{k=1}^\infty \sum_{\ell=1}^\infty \mathring{B}_{[j](k\ell)} \mathring{X}_{[j](i\ell)} \mathring{\Upsilon}_{k}(t) + \epsilon_i(t)\,, \qquad i = 1, \dots, n\,,
\end{equation*}
which in turn, by orthonormality of $\{\mathring{\Upsilon}_k:\,k\in\NN\}$, implies
\begin{equation}
    \small
    \mathring{Y}_{ik} = \sum_{j=1}^p \sum_{\ell=1}^\infty \mathring{B}_{[j](k\ell)} \mathring{X}_{[j](i\ell)} + \mathring{\varepsilon}_{ik}\,, \qquad i = 1, \dots, n\,, k \in \NN\,,
\end{equation}
where $\mathring{\varepsilon}_{ik} = \int_\mathcal{T} \epsilon_i(t) \mathring{\Upsilon}_k(t)\,dt$, $k\in\NN$. 
We truncate the expansions of functional responses after $K$ scores, and of all functional predictors after $L$ scores. The truncation introduces a first level of approximation error. A second level of approximation error comes from the fact that we do not directly observe the FPC eigenfunctions and their associated scores, but we estimate them from the sample at our disposal. Their empirical counterparts are ${\Upsilon}_k(t)$, ${\Xi}_{j\ell}(s)$, $Y_{ik}$, $X_{[j](i\ell)}$, and $B_{[j](k\ell)}$, $i=1,\dots,n$, $k=1,\dots,K$, $\ell=1,\dots,L$. As seen in Section \ref{subsec:coeff_representation}, if $\check{B}_{[j](kl)}$ is the minimizer of (\ref{eq:matrix_minimization_problem}), we 
have $\hat{\mathcal{B}}_j(t,s) = \sum_{k=1}^K \sum_{\ell=1}^L \check{B}_{[j](k\ell)} {\Upsilon}_k(t) {\Xi}_{j\ell}(s)$. 
 
Let $p_0$ denote the number of active features,
$C>1$ and $C_1 \leq C_2$
generic constants,
and $\Tilde{Z}_i = [ \mathring{X}_{[1](i1)}\xi_{11}^{-1/2}, \dots, \mathring{X}_{[p_0](iL)}\xi_{p_0L}^{-1/2} ]^T$ the $p_0L\times1$ vector
obtained by combining all active functional features of observation $i$. We 
assume the following regularity conditions:
\begin{enumerate}[label=\textbf{\Alph*.}]
    \item For any $i=1,\dots,n$, $j=1,\dots,p$ and
    $\ell \in\NN$, $\EE[\mathring{X}_{[j](i\ell)}^4] \leq C\mathring{\xi}_{j\ell}^2$, $\EE[\cnorm{\mathcal{X}_{ij}}^4] < \infty$, $\EE[\cnorm{\mathcal{Y}_i}^4] < \infty$, and $\EE[\cnorm{\epsilon_i}^2] \leq C$;
    
    \item The eigenvalues $\{\mathring{\upsilon}_k:\,k\in\NN\}$ and $\{\mathring{\xi}_{j\ell}:\,\ell\in\NN\}$ satisfy
    \vspace{-0.5cm}
    \begin{equation*}
    \begin{split}
        \small
        \mathring{\upsilon}_k \leq Ck^{-\alpha_1}\,, \quad \mathring{\upsilon}_k - \mathring{\upsilon}_{k+1} \geq C^{-1}k^{-\alpha_1-1}\,; \quad
        \mathring{\xi}_{j\ell} \leq C\ell^{-\alpha_2}\,, \quad \mathring{\xi}_{j\ell} - \mathring{\xi}_{j(\ell+1)} \geq C^{-1}\ell^{-\alpha_2-1}\,,
    \end{split}
    \vspace{-0.4cm}
    \end{equation*}
    for $k,\ell\in\NN$ and $j=1,\dots,p$, where $\alpha_1,\alpha_2 > 1$;
    
    \item $\lvert\mathring{B}_{[j](k\ell)}\rvert \leq Ck^{-\gamma_1}\ell^{-\gamma_2}$, for $k,\ell\in\NN$ and $j=1,\dots,p$, where $\gamma_1 > \alpha_1/2 +1$ and $\gamma_2 > \alpha_2/2 + 1$;
    
    \item $K,L,p_0\to\infty$, with $p_0 L^{\alpha_2/2 + 2}n^{-1/2} \to 0$, 
    $~(L^{2\alpha_2 + 2} + K^{2\alpha_1 + 2})/n \to 0$, \\
    $(K^{2\gamma_1 - 1} + L^{2\gamma_2 - 1})/n \to \infty$;
    
    \item $(\lambda_1 + \lambda_2)\omega_j 
    = o(1)$, $\ \max\{L^4p_0Kn^{-1},\,p_0KL^{-2\gamma_2-\alpha_2+1},\,K^3L^3n^{-1} \} = o(\lambda_1^2 
    )$, and \\ $\min_{j\in\{1,\dots,p_0\}} \lVert \mathring{B}_{[j]} \rVert_2 / 
    \lambda_1 
    \to \infty$ for $j=1,\dots,p$;
    
    \item $0 < C_1 \leq \rho_{min}(\EE[\Tilde{Z}_i\tilde{Z}_i^T]) \leq \rho_{max}(\EE[\Tilde{Z}_i\tilde{Z}_i^T]) \leq C_2 < \infty$, $i=1,\dots,n$, where $\rho_{min}(\cdot)$ and $\rho_{max}(\cdot)$ denote the smallest and the largest eigenvalues of a symmetric matrix, respectively;

    \item 
    $p=O_p(e^{n^\beta})$, with $\beta\in(0,1/2)$.
    
\end{enumerate}

\noindent 
Conditions \textbf{A}--\textbf{C} are standard in functional data analysis \citep[see, e.g.,][]{cai2006prediction, hall2007methodology, lei2014adaptive, imaizumi2018pca, cai2021variable, zhang2022subgroup}. Specifically,
\textbf{A} guarantees that the empirical covariance functions are consistent;
\textbf{B} ensures 
sufficient spacing between eigenvalues;
and \textbf{C} controls the tail behavior for large $k,\ell$. 
Conditions \textbf{D}--\textbf{F} are required for the consistency of estimators. 
Specifically, \textbf{D}
defines the right size for the truncation parameters $K$ and $L$
(large enough to capture most of the signal, but not as large as to include
increasingly noisy eigenfunctions);
\textbf{E} 
provides the rate for the tuning parameters $\lambda_1$ and $\lambda_2$, which are
critical for consistent estimation; and
\textbf{F} ensures invertibility of $\EE[\Tilde{Z}_i\Tilde{Z}_i^T]$. Finally, condition \textbf{G} allows the number of features to 
grow exponentially with the sample size $n$
-- a routine assumption in the high-dimensional setting \citep{buhlmann2011statistics, wainwright2019high}. 

To facilitate theoretical analysis, we follow \cite{kong2016partially, cai2021variable, zhang2022subgroup} 
and introduce the reparametrization $\Tilde{B}_{[j](k\ell)} = \mathring{B}_{[j](k\ell)}\xi_{j\ell}^{1/2}$. This guarantees that the FPC scores serving as
predictors have a common variability scale
(we note that the reparametrization is only useful for technical derivations;
we do not need it in the implementation of the \texttt{FAStEN} algorithm). The following Theorem provides theoretical guarantees 
for the proposed \texttt{FAStEN} estimator, showing that, asymptotically, it inherits oracle properties. A proof 
can be found in Appendix Section \ref{sec:supp_asympt}. 

\begin{theorem}
\label{th:oracle}
    Let $\tau^2 = K^3 +p_0(KL + nL^{-2\gamma_2 - \alpha_2})$. Under conditions \textbf{A}--\textbf{G}, there exists a local minimizer $\hat{B}$ of (\ref{eq:matrix_minimization_problem}) such that:
    \begin{enumerate}[label=(\roman*)]
        \item Selection consistency: $\PP[\hat{B}_{[j]} = 0,\, j=p_0+1,\dots,p] \to 1$;
        \item Estimation consistency: $\lVert \hat{B} - \Tilde{B} \rVert_2^2 = O_p(\tau^2/n)$.
    \end{enumerate}
\end{theorem}

\noindent 
In related literature, \cite{kong2016partially} study the behavior of FPC-based estimators in a mixed scalar-on-function regression setting with fixed number of functional predictors and diverging number of scalar predictors; \cite{cai2021variable} study estimation consistency of an FPC-based estimator in a function-on-function regression setting with fixed $p$; and \cite{zhang2022subgroup} study estimation and selection consistency of an FPC-based estimator in a scalar-on-function regression setting with diverging $p$. However, to the best of our knowledge, Theorem~\ref{th:oracle} is the first to prove selection consistency for an FPC-based estimator in a function-on-function regression setting with diverging $p$. 
The rate $\tau^2$ crucially depends on the assumption of sparsity, as is standard in high-dimensional statistics. In fact, the rate depends only on $p_0$ and not on $p$, making estimation possible \citep{wainwright2019high}. Furthermore, the rate provides theoretical suggestions on the choice of $K$ and $L$. The first two terms in $\tau^2$ imply that $K^3$ and $p_0KL$ must grow slower than $n$. Therefore, as highlighted in the previous section on the computational costs, it is crucial to choose $K$ and $L$ parsimoniously.

%
%

\section{Implementation specifics}
\label{sec:50shadows}

In this section, we discuss the various specifics of \texttt{FAStEN} implementation. Appendix Algorithm \ref{alg:implementation_detail} provides the corresponding pseudocode.


\paragraph{Path-search.}
We implement an efficient
path search to evaluate different values of the penalty parameter $\lambda_1$.
We compute 
solutions for a decreasing sequence of $\lambda_1$ values, starting from $\lambda_{max} = \max_j\lVert (X^TY)_{[j]}/\omega_j\rVert$, which selects $0$ active features. We use the solution obtained at the previous value of $\lambda_1$ to initialize the solver for the next step in the path (\textit{warm start}). At this stage, the 
$w_j$'s in 
\eqref{eq:matrix_minimization_problem} are all set to $1$, so that features are all weighted in the same fashion. We also allow the user to specify a maximum number of selected features; when this number is reached the path search is stopped, further reducing computation. 

To guide the choice of
$\lambda_1$ along the path, 
we consider two quantitative criteria; namely, the \textit{generalized cross-validation} (\textit{gcv}) and \textit{k-fold cross-validation} (\textit{cv}) mean squared errors. The former is obtained through the formula \citep{jansen2015generalized}
$gcv = \text{rss}(\check{B})/(n - K\nu)^2$, where 
$\text{rss}(\check{B})$ is the residual sum of squares, and $\nu = \text{tr} \big( X_\mathcal{J}  \big( X_\mathcal{J}^T X_\mathcal{J} + \la_2 I_{rK} \big) ^{-1} X_\mathcal{J}^T \big)$
are the degrees of freedom computed as in \cite{tibshirani2012degrees}.
This can be obtained from the original solution and is of course much less computationally expensive than the \textit{cv}, which requires that \texttt{FAStEN} be run multiple times (one per fold) for each value of $\lambda_1$. 
Before evaluating both criteria, we de-bias \texttt{FAStEN} estimates fitting a standard least squares on the selected features only. We indicate these estimates as $\check B^R$. This approach is known in the literature as \emph{relaxation} \citep{meinshausen2007relaxed, belloni2014debiasedols, zhao2021defense}.

Note
that the path search explores different 
values of $\lambda_1$ with $\lambda_2$ held constant. Multiple $\lambda_1$ path searches can be implemented, each corresponding to a different value of $\lambda_2$. In terms of selection, it is not necessary to explore a fine grid of $\lambda_2$ values, as its penalty does not induce sparsity and does not significantly impact the number of selected features.

\begin{figure}[!t]
    \hspace*{-1.5cm}
    \includegraphics[width=1.2\textwidth]{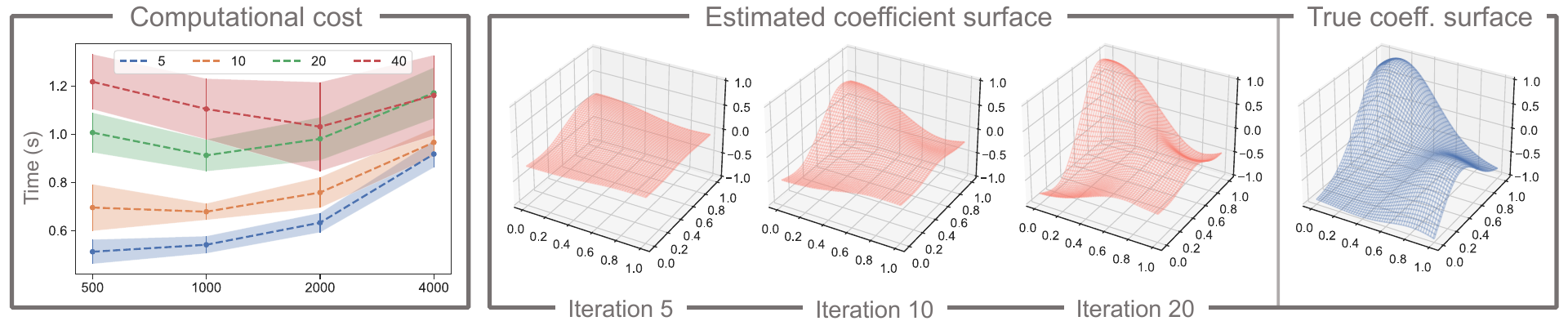}
    \vspace*{-0.7cm}
    \caption{
    Left panel: computational time comparison under different simulation scenarios for \texttt{FAStEN} with generalized cross validation ({\it gcv}) criterion and $K=3$.
    Simulations are run 10 times, with parameters $n=600$; $p = 500,1000,2000,4000$ (x-axis); $p_0=5,10,20,40$ (colors); $snr=1$ and ``difficult'' coefficients (see Section \ref{sec:sim_and_application}).
    Dashed lines join average computational times for various $n$'s and vertical bars represent $\pm 1$ standard deviation intervals across the 10 replications. Right panel: an instance of coefficient surface estimates at adaptive path search iterations 5, 10 and 20 (red), and true coefficient surface (blue). Simulation parameters are $n=600$; $p=8000$; $p_0=5$; $snr=10$ and ``easy'' coefficients. 
}
    \label{fig:complexity_adaptive}
\end{figure}
%

\paragraph{Adaptive implementation.}
The adaptive scheme starts from the features selected by the best solution of the path described above, i.e.~the path with all weights set to $1$. 
Next, we perform a new path search, this time with weights set to $\check \omega_j=1/\lVert \check{B}^R_{[j]}\rVert_2$.
To choose the best $\lambda_1$ in this adaptive path, the criteria are evaluated without relaxation; this is because using weights already mitigates the bias of the estimates.
\cite{zou2006adaptive, zou2009adaptive} first introduced adaptive paths for the Lasso and the Elastic Net problems, showing how weights can improve feature selection, possibly reducing the size of the active set and achieving the oracle property. Such improvements were also investigated by \cite{parodi2018simultaneous} in a function-on-scalar regression context, where the authors also demonstrated that weights can enhance coefficient estimation.
Similarly, through an extensive simulation study (see below), we show that in our proposal, the adaptive step improves the estimation of coefficient surfaces.
In the right panel of Figure \ref{fig:complexity_adaptive}, one can see how the quality of the estimates increases with the number of iterations along the adaptive path: considering different values of $\lambda_1$ combined with weights reduces estimation bias. 
Adding weights to the minimization in \eqref{eq:matrix_minimization_problem} is a significant novelty with respect to previous DAL implementations; it allows \texttt{FAStEN} to perform feature selection and achieve low-bias coefficient estimation simultaneously.

Since investigating a \emph{full adaptive} path may be computationally expensive, we also propose a \emph{soft adaptive} strategy which considers just one specific value of $\la_1$ (the one associated to the best solution of the unweighted path search). 
In this case, we set $\check \omega_j=\text{sd}_B/\lVert \check{B}^R_{[j]}\rVert_2$, where $\text{sd}_B$ is the standard deviation of  {\small $\left( \lVert\check{B}^R_{[1]}\rVert_2, \dots , \lVert\check{B}^R_{[p]}\rVert_2 \right)$}.
\paragraph{Selection of $K$ and $L$.}
The choice of $K$ and $L$ plays a crucial role in determining how accurately problem \eqref{eq:matrix_minimization_problem} approximates problem \eqref{eq:functional_minimization_problem}, as well as in
affecting the computational cost of the optimization procedure. Striking a balance between these factors is important for achieving both efficiency and accuracy. In many cases, since the FPCs are the most parsimonious orthonormal basis system, only a few components are sufficient to capture more than 90\% of the ${\mathbb{L}^2}$ norm of both the response $\mathcal{Y}$ and the predictors $\mathcal{X}$'s. Moreover, using small values of $K$ and $L$ allows us to focus on the most significant modes of variability, effectively screening out noise and reducing the dimensionality of the problem.

Even though \texttt{FAStEN} allows 
one to chose different numbers of basis functions for representing the response and the predictors, in our simulation study, we set $L = K$ to balance model complexity with predictive accuracy by minimizing the number of hyperparameters that need tuning. 
To tune $K$, we implement a greedy search.  We apply \texttt{FAStEN} starting with $K = 3$ (which explains more than 80\% of the response variability in all investigated scenarios). We incrementally increase $K$ and halt the search when the quantitative criterion (either $gcv$ or $cv$) stops decreasing. 

\paragraph{Supervised FPC reduction.}
The flexibility of \texttt{FAStEN} 
accommodates the use of different bases for 
$\mathcal{X}$'s and 
$\mathcal{Y}$. However, in the spirit of the work on supervised dimension reduction and envelope models \citep{li2022dimension, su2023envelope, zhang2024nonlinear}, one could represent all functional variables using the FPCs of the response $\mathcal Y$. 
Unlike standard FPC regression \citep{reiss2007functional},
where each functional feature $\mathcal X_j$ is represented
through its own principal components,
one can use the first $K$ FPCs of $\mathcal Y$ to 
represent all the features.
This approach 
substantially decreases the computational burden of computing FPCs, as the feature selection problem involves one response and potentially many features.
Moreover, this choice induces an implicit screening process, as features that are not well represented by the response FPCs are less likely to be highly predictive.
We refer to this implementation as the \textit{supervised}
version of \texttt{FAStEN} (where the reduced representation of all $\mathcal{X}$'s is guided by $\mathcal{Y}$), in contrast to the \textit{unsupervised} version (where the reduced representation of each 
$\mathcal{X}$
is guided by $\mathcal{X}$ itself).

%
%

\section{Simulation study}
\label{sec:sim_and_application}

In this section, we 
employ synthetic data to investigate performance and computational efficiency of \texttt{FAStEN}. 
First, we benchmark it against the method implemented in the \texttt{R package FRegSigCom} \citep{qi2018function}, which is the only competitor we found in the literature able to deal with high-dimensional scenarios ($p \gg n$).
Then, we compare the various \texttt{FAStEN} algorithmic implementations described in the previous section to assess their relative strengths and efficiencies.


\subsection{Simulation settings}

We generate the response additively according to model \eqref{eq:fof_problem}. Each feature $\mathcal{X}_{j}$, as well as the error $\epsilon$, are drawn from a $0$ mean Gaussian process with a Matern covariance function \citep{cressie1999classes} of the form
\begin{equation*}
\small
\begin{split}
        C(t,s) = \frac{\eta^2}{\Gamma(\nu)2^{\nu-1}}\Bigg(\frac{\sqrt{2\nu}}{l} \lvert t-s \rvert\Bigg)^\nu \chi_\nu\Bigg(\frac{\sqrt{2\nu}}{l} \lvert t-s \rvert \Bigg)
\end{split}
\end{equation*}
where $\chi_\nu$ is a modified Bessel function. For the $\mathcal X$'s we use point-wise variance $\eta^2=1$, range $l=0.25$ and smoothness parameter $\nu=3.5$. 
For $\epsilon$ we use $l=0.25$, $\nu=2.5$, and assume the variance $\eta^2_\epsilon$ to be the same across the entire time domain. The value of $\eta^2_\epsilon$ is set as a function of the \emph{signal-to-noise ratio (snr)} we define for different simulation scenarios. Specifically, we take $\eta^2_\epsilon = \text{var}(\mathcal Y_{true}))/
snr$, where $\text{var}(\mathcal Y_{true})$ is the global variance of the response $\mathcal Y$, and consider $snr$ = $1, 10, 100$. Scenarios are harder, from the perspective of both selection and estimation, the smaller their 
$snr$.

\begin{figure}[!t]
    \hspace*{-1.5cm}
    \includegraphics[width=1.2\textwidth]{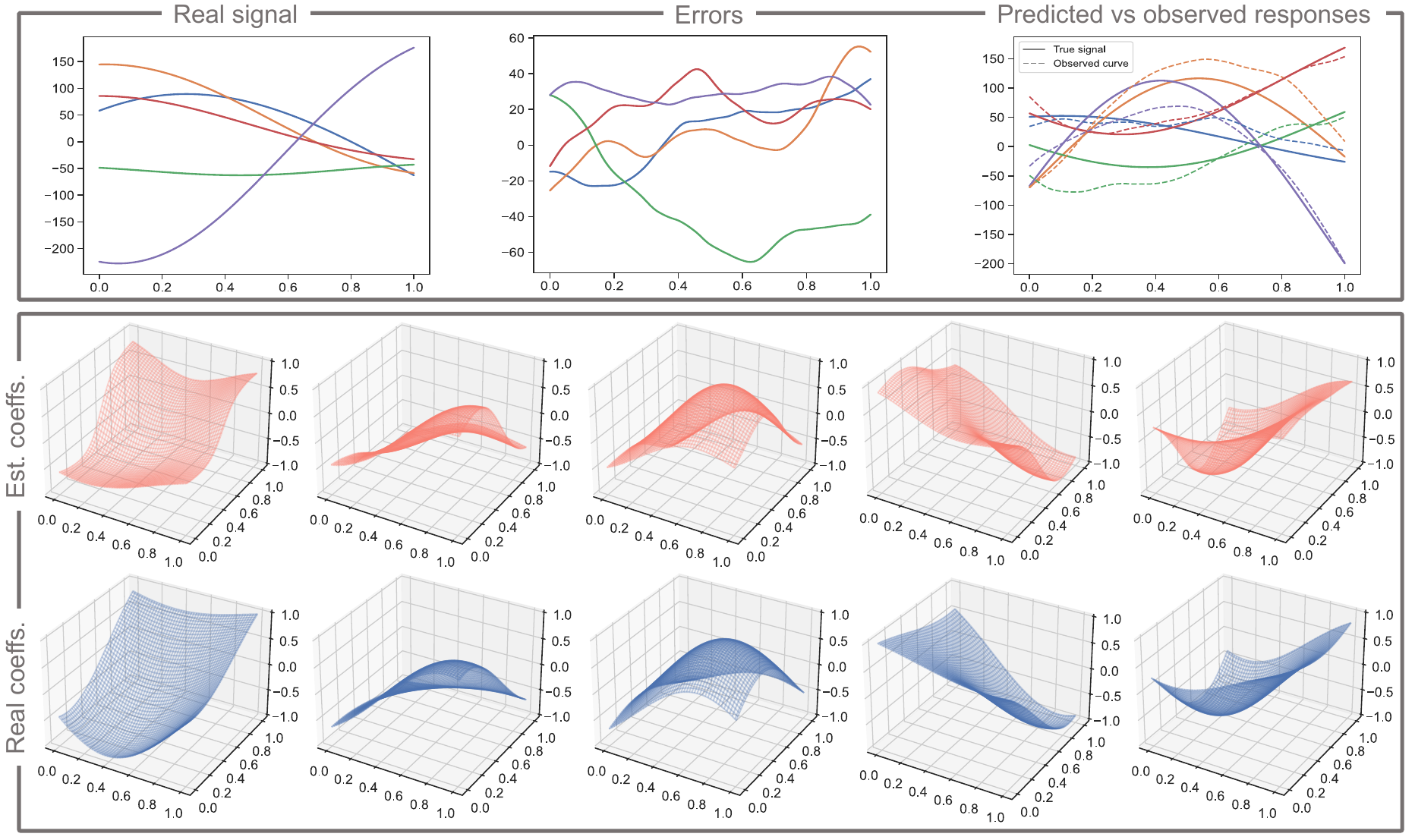}
    \vspace*{-0.5cm}
    \caption{ 
    Simulated data example (simulation parameters $n=600$; $p=8000$; $p_0=5$; $snr=10$ and ``easy'' coefficients).
    The top left panel shows a sample of 5 true signal curves
    (i.e.~responses before adding errors), the top center panel shows error curves (note the different vertical axis) and the top right panel shows the resulting response curves (true signal + error; dashed) along with the predictions obtained with \texttt{FAStEN} (solid).
    The bottom panel shows the 5 non-zero true coefficient surfaces (blue) and, above each, the corresponding \texttt{FAStEN} estimates (red). Both prediction and estimation performance appear strong in this illustration.
    }
    \label{fig:sim}
\end{figure}

For each scenario, $p_0$ indicates the number of active features, i.e.~non-zero regression coefficient surfaces. We generate these surfaces from a mixture of bivariate Gaussian distributions under two regimes of complexity. The \enquote{easy} regime is characterized by a single peak with standard deviation randomly drawn from a uniform distribution on $[0.2,0.3]$. The \enquote{difficult} regime is characterized by 2 or 3 peaks with standard deviations randomly drawn from a uniform on $[0.01,0.15]$.
Figure \ref{fig:sim} shows some instances of observed and predicted response curves, as well as true and estimated non-zero coefficient surfaces, for one specific scenario (the underlying simulation parameters are reported in the caption). 
In all scenarios, \texttt{FAStEN} is run with tolerances set to $10^{-6}$, starting from $\sigma^0=0.001$ and increasing by a factor of 5 at each iteration. 
Following standard practice in the literature \citep{friedman2010regularization, pedregosa2011scikit} we write $\lambda_1=\alpha c_\lambda\lambda_{max}$ and $\lambda_2=(1-\alpha)c_\lambda\lambda_{max}$, with $c_\lambda\in(0, 1]$ and $\alpha\in(0, 1)$. $c_\lambda$ determines the reduction with respect to $\lambda_{max}$, while $\alpha$ controls the relative weight of the two penalties. We set $\alpha=0.2$. For the line-search in \eqref{eq:linesearch}, we decrease the step size by a factor of 0.5 and set $\mu = 0.2$.

Before running \texttt{FAStEN} we standardize the response and each feature individually. For
each instance of 
each functional variable, say $\mathcal V$ for simplicity, we create a standardized version as $\left(\mathcal V(t) - \text{ave}(t) \right) / \text{sd}(t)$, where $\text{ave}(t)$ and $\text{sd}(t)$ are the point-wise average and standard deviation of all 
instances computed at 
$t$.

In each scenario, we evaluate performance 
with respect to \emph{selection, estimation, prediction,} and \emph{computational efficiency}. The numbers of 
false positives and false negatives 
track selection performance; the standardized mean square error of the coefficient surfaces ($\text{mse}(\hat{\mathcal B})$) measures the quality of estimation; the out-of-sample standardized mean square error of the responses ($\text{mse}_{out}(\hat{\mathcal Y})$) quantifies prediction performance; and CPU time in seconds measures computational efficiency. $\text{mse}(\hat{\mathcal B})$ and $\text{mse}_{out}(\hat{\mathcal Y})$ are defined as 
\begin{align*}
\small
\begin{split}
\label{eq:eval_criteria}
    \text{mse}(\hat{\mathcal B}) = \frac{1}{r} \sum_{j=1}^{r} \frac{\lVert\mathcal B_j - \hat{\mathcal B}_j\rVert_{\mathbb L^2 \times \mathbb L^2}}{\norm{\mathcal B_j}_{\mathbb L^2 \times \mathbb L^2}}
    \ \ \ , \ \ \
    \text{mse}_{out}(\hat{\mathcal Y}) = \frac{1}{n_{test}} \sum_{i=1}^{n_{test}} \frac{\lVert\mathcal Y_i - \hat{\mathcal Y}_i\rVert_{\mathbb L^2} }{\norm{\mathcal Y_i}_{\mathbb L^2}} 
    \ \ \ .
\end{split}
\end{align*}
Note that $\text{mse}(\hat{\mathcal B})$ is evaluated only on the selected features which belong to the true non-zero coefficients set. 
Concerning $\text{mse}_{out}(\hat{\mathcal Y})$, we evaluate it simulating a separate and independent \emph{test set} with size $n_{test} = n/3$ for each simulation run. We standardize both criteria to easily compare results across different simulation settings. 
\begin{figure*}[t!]
    \centering
    \hspace*{-1.7cm}
    \includegraphics[width=1.2\linewidth]{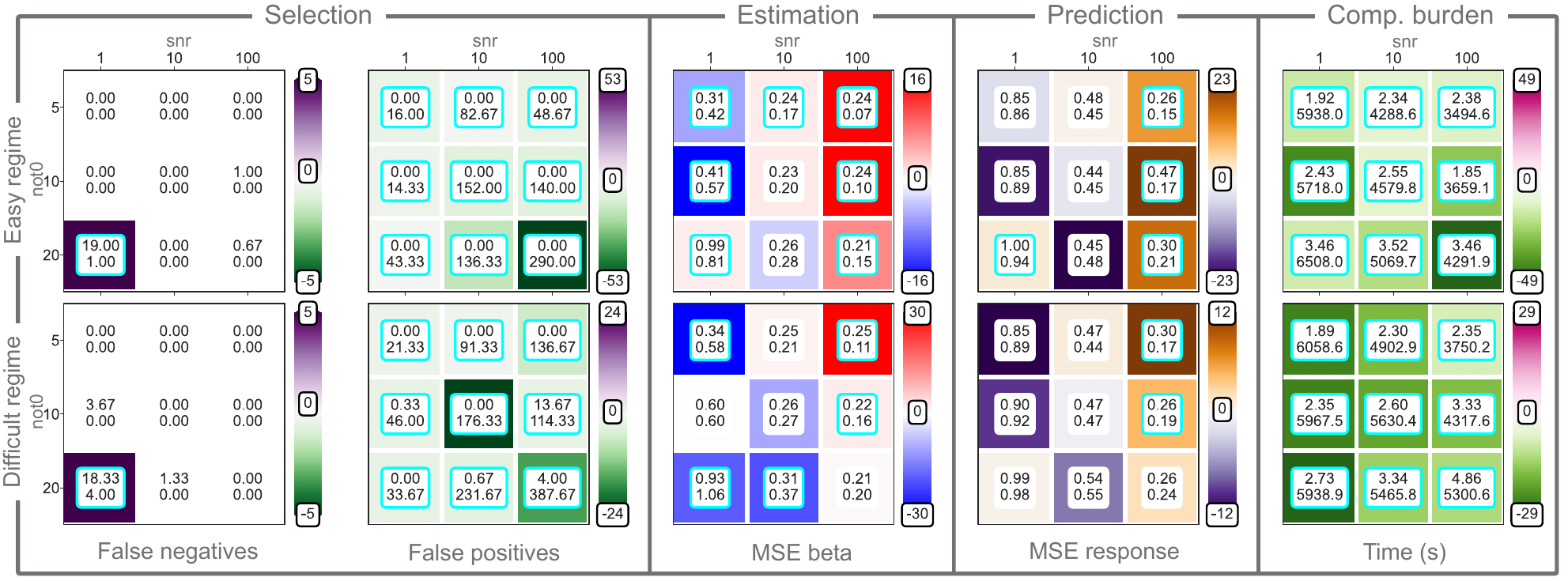}
    \vspace*{-0.5cm}
    \caption{Performance comparison between \texttt{FAStEN} with generalized cross validation (\textit{gcv}) as quantitative criterion, and \texttt{FRegSigCom}. Due to the computational burden of the latter, simulations are run only 3 times with $n=300$, $p=500$, different signal-to-noise ratios (1, 10, 100), number of active features (5, 10, 20) and coefficient complexity regimes (``easy'', ``difficult'').
    Each of the ten tables shows results for an evaluation metric (left to right; selection as false negatives and false positives, estimation error, prediction error and computational burden) and a coefficient complexity regime (top, ``easy''; bottom, ``difficult''). Within the cells of each table, the first and second rows contain average scores (across simulations) for \texttt{FAStEN} and \texttt{FRegSigCom}. Those scores are rimmed if the absolute value of their difference exceeds a threshold (5 for false negatives and false positives, 0.05 for estimation and prediction standardized MSEs, and 5 seconds for time).
    The background of each cell color-codes the difference between \texttt{FAStEN} and \texttt{FRegSigCom} average scores, divided by the standard deviation of the latter -- the color scale is provided on the right of each panel.
    Scales differ across panels, but in general a color associated to the negative part of the scale represents a better performance for \texttt{FAStEN}. 
    Estimation here is performed only on true positive features (we do not measure estimation quality on false positives).
    }
    \label{fig:cv_competitor}
\end{figure*}

\subsection{Simulation results}


\paragraph{\texttt{FAStEN} vs \texttt{FRegSigCom}.}
Results of the comparison 
between the two algorithms are
shown in Figure \ref{fig:cv_competitor}. Due to the computational demands of \texttt{FRegSigCom}, it is impractical to consider very high-dimensional settings. Therefore, our analysis is limited to scenarios with $n=300$ and $p=500$, and we perform only 3 simulation runs for each setting.
Appendix Figure \ref{fig:supp_competitor} shows additional results obtained from 10 replications for two specific scenarios, which are entirely consistent with those presented here. 

In terms of false negatives, the two methods perform similarly, with the exception of the 
noisiest and least sparse scenario ($snr=1$ and $p_0=20$), where \texttt{FAStEN} is much more conservative and misses the majority of the active predictors. 
In terms of false positives, \texttt{FAStEN} drastically outperforms \texttt{FRegSigCom} in every scenario, with an improvement in the range of two orders of magnitude. \texttt{FRegSigCom} does not meticulously explore the hyperparameter space and always selects a number of active features much larger than $p_0$.

In terms of estimation, \texttt{FAStEN} performs better in the hardest scenarios, characterized by low signal-to-noise ratio (especially so in the ``difficult'' coefficients regime), while \texttt{FRegSigCom} achieves better performances in high $snr$ scenarios; however, the absolute difference between the two average $\text{mse}(\hat{\mathcal B})$ is small.
Prediction performance follows a similar pattern: \texttt{FRegSigCom} performs better when the signal-to-noise ratio is high, while \texttt{FAStEN} demonstrates slightly superior performance in other scenarios.

Finally, \texttt{FAStEN} is significantly more computationally efficient than its competitor, achieving speedups of up to $\approx2500$ times in extreme cases (e.g.,~in the scenario with $snr=1$, $p_0=5$ and ``easy'' coefficients). This remarkable performance is achieved even while \texttt{FAStEN} performs an exhaustive grid search for $\lambda_1$, both regular and adaptive, and repeats the entire procedure for at least two different values of $K$.

Regarding $K$, out of the total 54 replications considered (3 replications per 18 different settings), \texttt{FAStEN} selects $K = 3$, $K = 4$, and $K = 5$ in 51, 2, and 1 cases, respectively.
This indicates that focusing on just a few principal components leads to better quantitative selection criteria. Furthermore, since a small $K$ is crucial for computational efficiency, these results further validate our methodology.

In conclusion, \texttt{FAStEN} drastically outperforms \texttt{FRegSigCom} in terms of selection and CPU time, improves the estimation of ``difficult'' coefficient surfaces in noisy instances, and is only slightly worse in terms of prediction when the problem is not very sparse and the signals are strong.


\paragraph{\texttt{FAStEN} algorithmic implementations.}
To investigate the various \texttt{FAStEN} implementations,
we focus on two scenarios: the \emph{``easy''} and \emph{``difficult''} regimes, with $p_0 = 10$, $snr = 10$, $n = 600$, and $p = 8000$. Notably, the efficiency of \texttt{FAStEN} enables exploration of significantly larger dimensional settings. In all comparisons, the baseline is the supervised version of \texttt{FAStEN} using \emph{gcv} as quantitative selection criterion and \emph{full adaptive} implementation.
\vspace{-0.3cm}
\\

\noindent 
\emph{Quantitative selection criteria.} Results comparing cross-validation (\textit{cv}) and generalized cross validation (\textit{gcv}) criteria are shown in Appendix Figure~\ref{fig:supp_gcv_cv}. The two criteria lead to nearly identical results in terms of selection, estimation, and prediction. However, as expected, $cv$ is more computationally expensive. 
\vspace{-0.3cm}
\\

\noindent 
\emph{Adaptive scheme.} Results comparing the \emph{full} and \emph{soft} adaptive implementations with a simpler version of \texttt{FAStEN} (where all $\omega$'s are set equal to 1) are 
shown in Appendix Figure~\ref{fig:supp_adaptive_soft_none}
The \emph{full} and \emph{soft} implementations produce similar results and significantly improve estimation and prediction performance compared to the non-adaptive version. This improvement comes at the cost of only a very small increase in computational time.
\vspace{-0.3cm}
\\

\noindent 
\emph{FPC reduction.} In Appendix Figure~\ref{fig:supp_X_Y},
we contrast the \emph{unsupervised} and \emph{supervised} versions of \texttt{FAStEN}. 
The two approaches yield nearly identical selection results. However, supervised \texttt{FAStEN} improves estimation quality, likely because it focuses on predictor components linked to the response. Additionally, as expected, the supervised version is faster because it computes the FPCs only for the response variable.

\section{AOMIC PIOP1 application}
\label{sec:application}
AOMIC PIOP1 comprises brain fMRI raw and pre-processed data collected during various task-based experiments on a sample of 216 university students \citep{snoek2021amsterdam}. We analyze data relative to the emotion anticipation task, which measures processes related to emotional anticipation and curiosity.
Specifically, 30 unique images were shown to each subject at regular intervals in a time span of 200 seconds. Each image was preceded by a cue. The cue could be valid, i.e.~describe correctly the subsequent image, or invalid, i.e.~not related to the image. 
The order of the images and the pairing with a valid or invalid cue was randomly selected for each subject. For all the subjects, $80\%$ of cues were valid -- see \citep{snoek2021amsterdam} for more details.

\begin{figure}[t]
    \hspace*{-1.5cm}
    \includegraphics[width=1.2\linewidth]{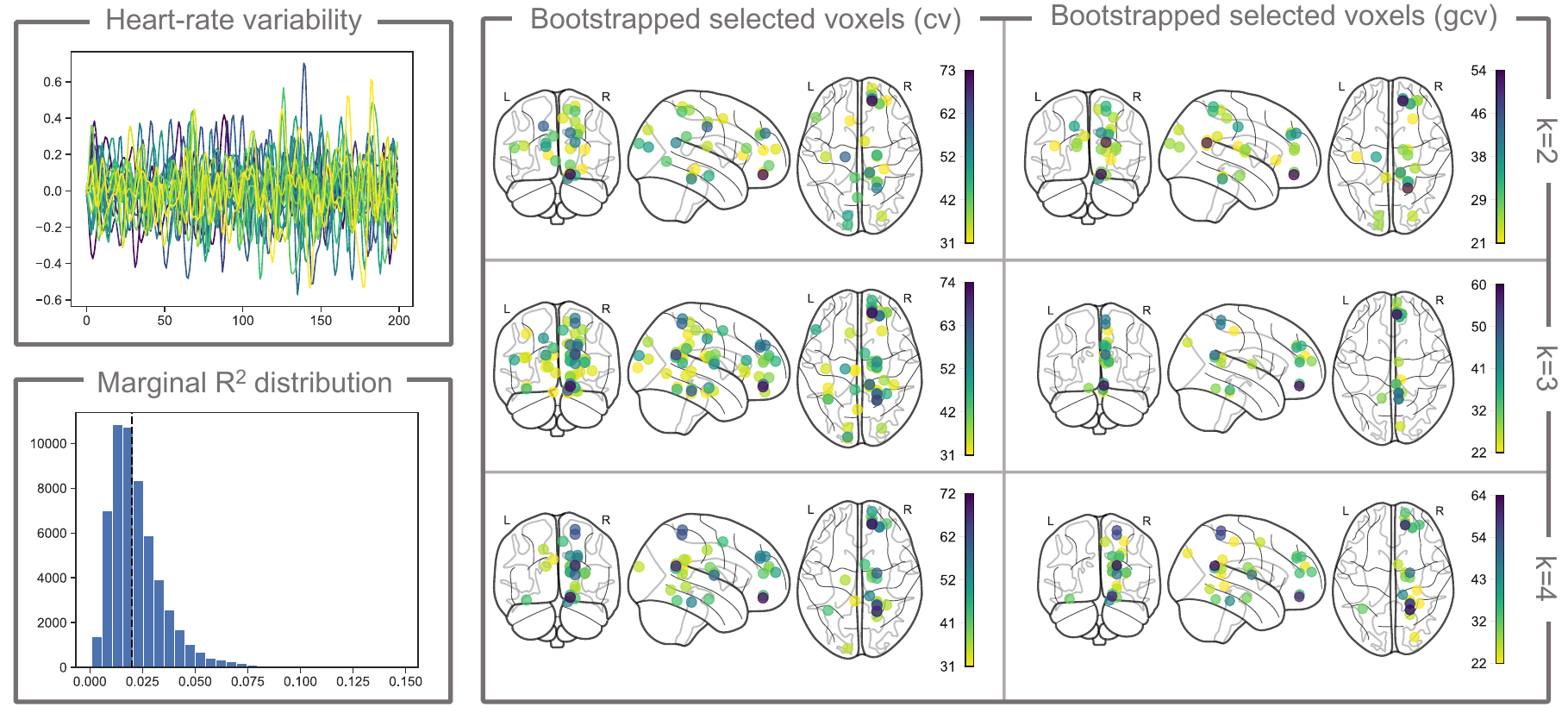}
    \vspace{-0.6cm}
    \caption{ 
    AOMIC PIOP1 data.
    Top left: heart rate curves (used as response) for a selection of 20 subjects in the study.
    Bottom left: histogram of $R^2$s from the marginal regressions of heart rate on each of the voxels. The threshold at 0.02 (dashed vertical line) is used to screen voxels. 
    Right panel: voxels selected at least 30 times applying FAStEN to 100 bootstrap samples from the original data, using the \textit{cv} (left) or \textit{gcv} (right) 
    criteria and $L= K = 2, 3, 4$ (top, centre, bottom, respectively). 
    For each 3D coordinate in the brain, the voxels are projected on (from left to right) the Coronal, Sagittal, and Horixontal axis.
    Dot colors represent frequency of selection (out of 100, see color scales).
    }
    \label{fig:brain_app}
\end{figure}

In addition to activation at each voxel (i.e., roughly, location of the brain in 3D coordinates), various measurements were taken on the subjects along the 200 seconds duration of the experiment. We focus on heart-rate variability (HRV), which is a convolution of the cardiac trace with a cardiac response function \citep{chang2009influence}; 
a selection of 20 HRV curves is plotted in the top left panel of Figure \ref{fig:brain_app}.
Our analysis aims to understand which are the voxels (features) whose activation is associated with HRV (response). 
We overlap brain masks of different subjects in order to get an homogeneous brain model, also removing scanned space outside the brain. 
This leads to a data set of 188 subjects and 55,551 voxels, whose activity is evaluated at $200$ time points. 
Curves are smoothed using a smoothing spline, i.e. cubic \textit{B-splines} with knots at each time point and a roughness penalty on the second derivative \citep{ramsay2005}. For each voxel activity, and for HRV, the smoothing parameter is selected by minimizing the average generalized cross-validation error \citep{craven1978smoothing} across the curves of the 188 individuals. These pre-processing steps are performed with the \texttt{R package fda} \citep{ramsayfdapackage}. 

We employ the \emph{supervised} version of \texttt{FAStEN} due to the high number of predictors, utilizing the \emph{full} adaptive strategy and both quantitative selection criteria (\textit{cv} and \textit{gcv}). 
Thanks to the computational efficiency of \texttt{FAStEN}, we are able to run it on the entire dataset, comprising 55,551 voxels. To the best of our knowledge, data of this size has not been explored through function-on-function regression to date. Notably, the number of selected voxels remains rather stable as we vary $K=L$ and the selection criterion. $K=3$ minimizes both \emph{cv} and \emph{gcv}.
To gauge stability, we also perform a different analysis in which \texttt{FAStEN} is preceded by a preliminary screen of the features.
We marginally regress HRV against the activation of individual voxels, and we identify the 26,682 voxels 
for which the marginal regression has a coefficient of determination $R^2>0.02$ 
(the histogram of such $R^2$'s is displayed in the bottom left panel of Figure \ref{fig:brain_app}). 
Running \texttt{FAStEN} (with \textit{gcv} and $L=K=3$) on the full and on the screened set of features produces exactly the same selection of 48 active voxels.
This suggests that, even in complex real data sets, \texttt{FAStEN} can indeed pinpoint relevant signals, parsimoniously and stably, regardless of the number of irrelevant signals among which they are interspersed. 
The $R^2$ of the model comprising 48 predictors is $0.12$. Notably, this is similar to $R^2$ values obtained in related literature -- see \citet{bailes2023resting}. We also note that, with these 48 selected predictors, one could fit a linear function-on-function model and increase the number of parameters for surface estimation. This would reduce the in-sample prediction error but would not affect the out-of-sample prediction error (see Appendix Figure~\ref{fig:supp_app_r2}).

Finally, and again 
in pursuit of a stability perspective, we use the 26,682 screened voxels to run \texttt{FAStEN} on 100 bootstrap samples of individuals, 
varying $K$ and $L$ and the selection criterion, and counting the number of times each voxel is selected. 
The right panel of Figure \ref{fig:brain_app}
shows the voxels selected at least 30 times (out of 100) by \texttt{FAStEN} with $L=K=2,3,4$ and considering \textit{cv} and \textit{gcv} 
($\alpha$ is fixed to $0.2$).
%
We obtain remarkably consistent results across values of $K$ and $L$ and selection criteria -- with $gcv$ being slightly more conservative than $cv$. Specifically, two areas of persistent association with the response appear to emerge. The first area 
is the mesencephalon
(shown in the center of the Horizontal axis in Figure \ref{fig:brain_app})
which is critical to the regulation of vital tasks, such as respiration and heart rate control. The second area is the orbitofrontal cortex (shown in the upper right of the Horizontal axis in Figure \ref{fig:brain_app})
which is involved in the reward and punishment system, and may thus be associated with emotion anticipation. Hence, while drawing strong domain conclusion is beyond the scope of this paper, we do have some evidence that the brain locations pinpointed by our analysis may indeed be related to heart rate variability patterns during the emotion anticipation task under consideration \citep{rolls2008orbitofrontal, holstege2014periaqueductal}.

\section{Conclusions}
\label{sec:conclusion}

In this paper we proposed \texttt{FAStEN}, a very efficient methodology to simultaneously perform feature selection and estimation in functional regression. 
While we focused on the most complex context, i.e.~the function-on-function case, 
additional information provided in the Appendix (scalar-on-function case) and prior work from \cite{boschi2021highly} (function-on scalar case) cover the whole spectrum of regressions comprising functional data. Function-on-function feature selection has not been extensively studied to date, perhaps because of the challenges posed by the complexity and the dimension of the problem. 
\texttt{FAStEN} exploits the properties of FPC -- which allow us to represent the functions in a finite space capturing their dominant modes of variability. 
\texttt{FAStEN} also exploits a novel adaptive version of the DAL to minimize a 
newly defined, Elastic Net type objective function comprising two different penalties.
The DAL algorithm is very efficient in sparse scenarios, where it leverages the structure of the Augmented Lagrangian second order derivative to significantly reduce the dimension of the problem and the computational cost. 
Finally, \texttt{FAStEN} comprises an adaptive scheme which can be used to enhance estimation performance. 

We established asymptotic properties of the \texttt{FAStEN} estimator, proving 
oracle estimation and selection consistency. The theory we developed holds in a general setting and provides 
guarantees on the behavior of FPCs in function-on-function regressions where the number of features is allowed to diverge with the sample size.

Our simulation study demonstrates that \texttt{FAStEN} leads to dramatic improvements in feature selection and computational time, without compromising estimation or prediction accuracy.
We also applied \texttt{FAStEN} to a brain fMRI data set, exploring more than 55,000 voxel activation curves for potential associations with variation in heart rates. Thanks to \texttt{FAStEN} computational efficiency, we were in a position to perform analyses involving multiple runs of the procedure (e.g.,~with different parameter specifications and on numerous bootstrap samples) and thus reach stable and enlightening results.
These outline an association between two well identified brain regions and heart rate variability during emotion anticipation.  

We plan to further expand our approach by 
investigating more complex penalties and incorporating both scalar and functional features in the same selection problem.
Other interesting developments concern the choice of basis system and the number of components ($K$ and $L$) used to approximate functional observations.
Formulating a quantitative criterion to determine a different and specific $K$ or $L$ for each variable and for each task (selection and estimation) would allow one to represent each feature more accurately and may improve the overall performance of the algorithm. Also, we plan to expand our theory in order to incorporate the group structure that may naturally arise within functional predictors (e.g. voxel activation curves recorded in the same region of the brain may be grouped together). 

Finally, one could investigate the properties of the estimated coefficient surfaces. 
While this work focuses on deriving theoretical guarantees in the reduced representation space via functional principal components, future work could explore convergence in the original functional space. 
Moreover, providing valid confidence bands with global coverage for these surfaces is non-trivial, even within the linear functional regression framework, due to the infinite-dimensional nature of the problem and the complex dependencies between functions. 
Although local normal approximation of the estimates can be guaranteed, there are no assurances for the global distribution, making the estimation of simultaneous confidence bands very challenging.
This is further compounded in penalized settings, such as in the classical LASSO, where the selection of regularization parameters introduces an additional bias component to the estimates.

%
%


\section*{Acknowledgements}

\noindent The work of Francesca Chiaromonte is partially supported by the Huck Institutes of the Life Sciences at Penn State and by the Sant'Anna School of Advanced Studies; the work of Matthew Reimherr is partially supported by the Grant NSF SES-1853209. Tobia Boschi gratefully acknowledges Dr. Ludovica Delpopolo Carciopolo for
useful discussions about the development and the coding implementation of the optimization algorithm.


%
%

\clearpage
\bibliography{bib}

%
%
\newpage
\appendix

\renewcommand{\thefigure}{\thesection.\arabic{figure}}
\setcounter{figure}{0}    

\section{Algorithmic results}
\label{sec:supp_proofs}

\subsection{Proposition \ref{prop:p_star}}
\label{subsec:proof_prop1}

 \noindent \textbf{\textit{Proof}}
First, note that $\pi(B) = \sum_{j=1}^p \pi (B_{[j]})$ is a separable sum: from \citet[Sec.~3.3.1]{boyd2004convex} we have $\pi^*(Z) = \sum_{j=1}^p \pi^* (Z_{[j]})$. Then, note that $\norm{B_{[j]}}_2 = \norm{\text{vec}(B_{[j]})}_2$, where $\text{vec}(B_{[j]}) \in \mathbb{R}^{KL}$ is the vector obtained by stacking all the elements of $B_{[j]}$ in one column. 
Since $\pi (B_{[j]})$ depends just on $\norm{B_{[j]}}_2$, it is possible to compute its conjugate function starting from $\text{vec}(B_{[j]})$. Now, we can use the result in \emph{Proposition 1} of \cite{boschi2021highly}, where the authors compute the elastic-net penalty conjugate function for a vector of dimension $K$. Following their proof, and incorporating the adaptive weights $\omega$'s in the penalty parameters $\la_1$ and $\la_2$, we obtain:
\begin{equation*}
\small
	\pi^*(Z) =  \sum_{j=1}^p (2\omega_j\la_2)^{-1}
		\begin{cases}
			\big(\norm{\text{vec}(Z_{[j]})}_2 - \omega_j\la_1\big)^2  & \norm{\text{vec}(Z_{[j]})}_2 > \omega_j\la_1 \\
			0  & o.w.
		\end{cases} \ .
\end{equation*}
This concludes the proof since, once again, we have $\norm{\text{vec}(Z_{[j]})}_2 = \norm{Z_{[j]}}_2$.

\subsection{Proposition \ref{prop:z_update}}
\label{subsec:proof_prop2}

 \noindent \textbf{\textit{Proof}}
$\pi(B) = \sum_{j=1}^p \pi (B_{[j]})$ is a separable sum, then 
$\prox_{\sa \pi}(B) =  \big(\prox_{\sa \pi}(B_{[1]}), 	\dots, \prox_{\sa \pi}(B_{[p]}) \big)^T$ \citep{beck2017first} (Remark~6.7).
For a generic vector $a$, we have $\prox_{\sa \la_1 \norm{\cdot}_2}(a) = \big[ 1 - \norm{a}_2^{-1} \sa \la_1 \big]_+ a$ \citep{fan2016high}, and $\prox_{(\sa \la_2/2)\norm{\cdot}_2^2}(a) = (1 + \sa \la_2)^{-1}a$ \citep{beck2017first} (6.2.3). Following the same reasoning of the previous proof, it is possible to extend these results to a generic matrix $A$ and including the weights $\omega$'s in the penalty parameters $\la_1$ and $\la_2$.
Next, note $\norm{\cdot}_2^2$ is a proper closed and convex function. Thus, we can compose $\prox_{(\sa \omega_j \la_2/2)\norm{\cdot}_2^2}$ and $\prox_{\sa \omega_j \la_1\norm{\cdot}_2}$ as described in \citet{parikh2014proximal}, obtaining:
\begin{equation*}
\small
	\prox_{\sigma \pi}(B_{[j]}) =  (1 + \sa \omega_j \la_2)^{-1} \left[1- \norm{B_{[j]}}_2^{-1} \sa \omega_j \la_1  \right]_{+} B_{[j]}  \ .	
\end{equation*}
Next, we prove $\bar Z =  \prox_{\pi^*/\sa} \big( B/\sa - X^T \bar V \big)$.
Computing the derivative of $\mathcal{L}_\sa \left(\bar Z, \bar V, B \right)$ as defined in \eqref{eq:augmented_lagrangian} with respect to $Z_{[j]}$, and setting it equal to 0, we have: 
\begin{equation*}
\small
	B_{[j]} / \sa - \bar V^T X_{[j]} - \bar Z_{[j]} = \nabla \pi^*(\bar Z_{[j]}) / \sa
\end{equation*}
Setting $t = B_{[j]} / \sa - \bar V^T X_{[j]} $, $u = \bar Z_{[j]}$, and $f = \pi^* / \sa$, we have $t - u = \nabla f(u)$. By the sub-gradient proximal operators characterization \citep{correa1992subcharacterization}, we can state $u = \prox_f(t)$, i.e.: $\bar Z_{[j]} =  \prox_{\pi^*/\sa} \big( B_{[j]}/\sa - X_{[j]}^T \bar V \big)$. Next, note $\bar Z = (\bar Z_{[1]}, \dots, \bar Z_{[p]})^T$.

The last equality of \emph{Proposition 2} follows by the \emph{Moreau decomposition}. Indeed, for a generic matrix $A$, we have:
$A = \prox_{\sa p}(A) + \sa \prox_{\pi^*/\sa}(A/ \sa)$, $\sa > 0$.

\subsection{Theorem \ref{th:v_update}}
\label{subsec:proof_theorem1}

 \noindent \textbf{\textit{Proof}}
Recall $T=B-\sa X^T V$, which means $T_{[j]} = B_{[j]} - \sa V^T X_{[j]}$, and $\Bar{Z} = T/\sa - \prox_{\sa p}\left(T \right)/\sa$. \\

\noindent 
\textbf{\emph{(i)}}~~ We need to compute $\psi(V) = \mathcal{L}_\sigma (V\,|\,\Bar{Z},B)$.
Plugging $\Bar{Z}$ in \eqref{eq:augmented_lagrangian}, we get: 
\begin{align}
    \small
    \label{eq:psi_medium}
    \begin{split}
        \psi(V) &= h^*(V) + \pi^*(\bar Z) - 
        \frac{1}{\sa} \sum_{j=1}^p \langle B_{[j]}, B_{[j]} \rangle 
        +  \frac{1}{\sa} \sum_{j=1}^p \langle B_{[j]}, \prox_{\sa \pi}\big( T_{[J]} \big) \rangle +
        \frac{1}{2\sa} \sum_{j=1}^p \lVert B_{[j]} -\prox_{\sa \pi}\big( T_{[J]}\big) \rVert_2^2 = \\
        & = h^*(V) + \pi^*(\bar Z) + \frac{1}{2 \sa} \sum_{j=1}^p \normbig{\prox_{\sa \pi}\big(T_{[j]} \big)}_2^2 - \frac{1}{2\sa} \sum_{j=1}^p\norm{B_{[j]}}_2^2 \ .
    \end{split}
\end{align}
Next, we compute $\pi^*(\bar Z) = \sum_{j=1}^p \pi^* (\bar Z_{[j]})$. Note, the explicit form of $\bar Z_{[j]}$ is:
\begin{equation*}
\small
	\bar Z_{[j]} = T_{[j]}/\sa - \prox_{\sa \pi} \big(T_{[j]} \big) /\sa =  
		\begin{cases}
			(1 + \sa \omega_j \la_2)^{-1} \big(\omega_j \la_2 + \norm{T_{[j]}}_2^{-1} \omega_j \la_1 \big) T_{[j]} & \norm{T_{[j]}}_2 > \sa \omega_j \la_1 \\
			T_{[j]}/\sa & o.w.
		\end{cases} \ .
\end{equation*}
We focus on the case $\norm{\bar Z_{[j]}}_2 > \omega_j\la_1$, i.e. where $\pi^* (\bar Z_{[j]}) \neq 0$. Note, this implies $\norm{ T_{[j]}}_2 > \sa \omega_j\la_1$, i.e.~$\prox_{\sa \pi} \big(T_{[j]} \big) \neq 0$. Plugging the explicit form of $\bar Z_{[j]}$ in the definition of $\pi^*$ given in \eqref{eq:p_star}, we have:
\begin{align*}
\small
    \begin{split}
        \pi^* (\bar Z_{[j]}) & = (2\omega_j\la_2)^{-1} \left((1 + \sa \omega_j \la_2)^{-1} \big(\omega_j \la_2 + \norm{T_{[j]}}_2^{-1} \omega_j \la_1 \big) \normbig{T_{[j]}}_2 - \omega_j\la_1 \right)^2 = \\
        & = (\omega_j\la_2 / 2) \left((1 + \sa \omega_j \la_2)^{-1} \left(\norm{ T_{[j]}}_2 - \sa \omega_j \la_1 \right) \right)^2 
    \end{split}
\end{align*}
Furthermore, starting from the definition of $\prox_{\sa \pi}$ given in \eqref{eq:prox_op}, we have: 
\begin{align*}
\small
    \begin{split}
        \norm{\prox_{\sa \pi}\left(T_{[j]}\right)}_2^2 & = \left( (1 + \sa \omega_j \la_2)^{-1} \left(1- \norm{T_{[j]}}_2^{-1} \sa \omega_j \la_1 \right) \norm{T_{[j]}}_2 \right)^2 = \\
        & = \left((1 + \sa \omega_j \la_2)^{-1} \left(\norm{ T_{[j]}}_2 - \sa \omega_j \la_1 \right) \right)^2
    \end{split}
\end{align*}
Thus, we have $\pi^* (\bar Z_{[j]}) = (\omega_j\la_2 / 2) \norm{\prox_{\sa \pi}\left(T_{[j]}\right)}_2^2$, which leads to \\
$\pi^* (\bar Z) = \sum_{j=1}^p (\omega_j\la_2 / 2) \norm{\prox_{\sa \pi}\left(T_{[j]}\right)}_2^2$. Plugging $\pi^* (\bar Z)$ in \eqref{eq:psi_medium} concludes the proof. \\

\noindent 
\textbf{\emph{(ii)}}~~ We have $\nabla h^*(V) = Y + V$. Thus, to prove \emph{(ii)}, it is sufficient to show: 
{\small 
$$\nabla_V \left(\frac{1}{2\sa} \sum_{j=1}^p \left( 1 + \sa \omega_j \la_2 \right) \normbig{\prox_{\sa \pi}\big(T_{[j]}\big) }_2^2\right) = - X \prox_{\sa \pi}(T)$$
}
Before proceeding note that: 
\begin{equation*}
    \nicefrac{\partial T_{[j]}}{\partial V} = -\sa X_{[j]} \text{~~and~~} \nicefrac{\partial \prox_{\sa \pi}\left(T_{[j]}\right)}{\partial T_{[j]}} =  (1 + \sa \omega_j \la_2)^{-1} 
\end{equation*}
The second equality holds because $\nicefrac{\partial \prox_{\sa \pi}\left(\norm{T_{[j]}}_2^{-1} T_{[j]}\right)}{\partial T_{[j]}} = 0$. We can now apply the chain rule for derivatives: 
\begin{align*}
\small
    \begin{split}
        \nabla_V & \left(\frac{1}{2\sa} \sum_{j=1}^p \left( 1 + \sa \omega_j \la_2 \right) \normbig{\prox_{\sa \pi}\big(T_{[j]}\big) }_2^2\right) = 
        \\
        & = \frac{1}{\sa} \sum_{j=1}^p \left( 1 + \sa \omega_j \la_2 \right)
        \nicefrac{\partial \norm{\prox_{\sa \pi}\left(T_{[j]}\right)}_2^2}{\partial \prox_{\sa \pi}\left(T_{[j]}\right)} \cdot
        \nicefrac{\partial \prox_{\sa \pi}\left(T_{[j]}\right)}{\partial T_{[j]}}\cdot
        \nicefrac{\partial T_{[j]}}{\partial V} = \\
        & = - \sum_{j=1}^p X_{[j]} \prox_{\sa \pi}\left(T_{[j]}\right) = 
        - X \prox_{\sa \pi}\left(T\right)
    \end{split}
\end{align*}
where the last equality holds because $X$ and $T$ are both block matrices. \\

\noindent 
\textbf{\emph{(iii)}}~~ To prove \emph{(iii)} we need to find $H_\psi (V)$. We can compute it as follows: 
{\small
$$ H_\psi (V) = \nabla_V\big( \nabla \psi (V) \big) = I_{nK} + \nabla_V \big(-X \prox_{\sa \pi} (T) \big).$$
}
We just need to show $\nabla_V \big(-X \prox_{\sa \pi} (T) \big) = \sa \hat X_\mathcal{J} Q_\mathcal{J} \hat X_\mathcal{J}^T$. Before proceeding, note that: 
\begin{equation*}
\small
    \nabla_{\prox_\sa \pi (T)}\big(X \prox_{\sa \pi} (T) \big) = \left( X_\mathcal{J} \otimes_{Kron} I_K \right) \text{~~and~~} \nabla_{V}\big(X^T V\big) = \left( X \otimes_{Kron} I_K \right)^T.
\end{equation*}
Both equalities are a special case of the result $T4.6$ in \cite{brewer1978kronecker}. Applying the chain rule of derivative, we obtain:
\begin{align*}
\small
    \begin{split}
       \nabla_V \big(-X \prox_{\sa \pi} (T) \big) & = - \nabla_{\prox_\sa \pi (T)}\big(X \prox_{\sa \pi} (T) \big) \cdot \nabla_{T} \big(\prox_{\sa \pi} (T) \big) \cdot
       \nabla_{V} \big(T\big) = \\
       & = \sa \left( X_\mathcal{J} \otimes_{Kron} I_K \right) \nabla_{T} \big(\prox_{\sa \pi} (T) \big) \left( X_\mathcal{J} \otimes_{Kron} I_K \right)^T.
    \end{split}
\end{align*}
Note $\nabla_{T} \big(\prox_{\sa \pi} (T) \big) \in \mathbb{R}^{pKL \times pKL}$ is a block diagonal matrix. Indeed, the blocks, each of dimension $KL \times KL$, are given by $ \nabla_{T_{[s]}} \left(\prox_{\sa \pi}\left(T_{[j]} \right)\right)$ for $s,j = 1, \dots, p$ and are $0$ if $s\neq j$. \\

Consider the case $s=j=1$. If $\norm{T_{[1]}}_2 \leq \sa\omega_1\la_1$, then $ \nabla_{T_{[1]}} \left(\prox_{\sa \pi}\left(T_{[1]} \right)\right) = 0$. Otherwise, let $\vec T_1 = \text{vec}\left( T_{[1]} \right)$ be the gradient vectorization with elements $t_i, \dots, t_{KL}$. Moreover, note that $\frac{\partial \norm{T_{[1]}}_2} {\partial t_i} = \norm{T_{[1]}}_2^{-1} t_i$.
Then, the element in position $(i,j)$ of $\nabla_{\vec T_1} \big(\prox_{\sa \pi}\big(\vec T_1 \big)\big)$ is given by:   
\begin{equation*}
\small
\label{eq:delta_prox}
	\frac{ \partial \prox_{\sa \pi}(t_i)} {\partial t_j} = (1 + \sa \omega 1 \la_2)^{-1}
	\begin{cases}
		  1 - \sa \omega_1 \la_1 \norm{T_{[1]}}_2^{-1} + \sa \omega_1 \la_1 \norm{T_{[1]}}_2^{-3}  t_i^2   &  i = j \\
		  \sa \omega_1 \la_1 \norm{T_{[1]}}_2^{-3} t_i t_j &  i \ne j
	\end{cases} \ .
\end{equation*}
The last equation shows $\nabla_{T_{[1]}} \left(\prox_{\sa \pi}\left(T_{[1]} \right)\right) = P_{[1]}$. Without loss of generality, we have $\nabla_{T_{[j]}} \left(\prox_{\sa \pi}\left(T_{[j]} \right)\right) = P_{[j]}$, for $j = 1,\dots, p$.
Let $Q$ be the $pKL \times pKL$ block diagonal matrix such that each block $Q_{[j]}$ is equal to $P_{[j]}$ if $j \in \mathcal{J} = \big\{ j~:~ \lVert T_{[j]} \rVert_2  \ge \sa \omega_j \la_1 \big\}$ and $0$ otherwise. We have $\nabla_{T} \big(\prox_{\sa \pi} (T) \big) = Q$ and
{\small
$$ H_{\psi}(V) = I_{nK} + \sa \left( X \otimes_{Kron} I_K \right) 
Q
\left( X \otimes_{Kron} I_K \right)^T \ . $$}
Finally, we just note that given the form of $Q$ only the blocks in $\mathcal{J}$ are selected and involved in the matrix multiplication; therefore: 
{\small
$$
\sa \left( X \otimes_{Kron} I_K \right) 
Q
\left( X \otimes_{Kron} I_K \right)^T = \sa \hat X_\mathcal{J} Q_\mathcal{J} \hat X_\mathcal{J}^T \ .
$$
}

%
%
\section{Extension to scalar-on-function regression}
\label{sec:supp_scalar}

The scalar-on-function regression model \citep{ramsay2005, kokoszka2017introduction} describes a scenario where the response is a scalar variable, while the features are curves.
Using the same notation introduced in Section \ref{sec:intro}, 
the scalar-on-function model with $n$ observations and $p$ features takes the following form:
\begin{align}
\small
\label{eq:scalar_model}
\begin{split}
    Y_i = \sum_{j=1}^p \int_{\mathcal{S}} \mathcal{B}_j(s)\mathcal{X}_{ij}(s)\,ds + \epsilon_i \qquad i = 1, \dots, n \ .
\end{split}
\end{align}
The model assumes that each feature $\mathcal{X}_j$ can affect the response vector $Y \in \mathbb{R}^n$ in a different way along the time domain. The coefficients $\mathcal{B}_j$ are curves and not surfaces as in the function-on-function framework. 
The \texttt{FAStEN} optimization problem has the form
\begin{align}
\small
\begin{split}
    \label{eq:functional_minimization_problem_scalar}
    \min_{\mathcal{B}_1, \dots, \mathcal{B}_p} ~ \Bigg[ 
    \frac{1}{2} \Big\lVert~Y-\sum_{j=1}^p \int_{\mathcal{S}} \mathcal{B}_j(s)\mathcal{X}_{j}(s)\,ds~\Big\rVert^2_2 +
    \sum_{j=1}^{p} \omega_j \left(
     \lambda_1 \lVert \mathcal{B}_j\rVert_{\mathbb{L}^2}+ 
     \frac{\lambda_2}{2}\lVert \mathcal{B}_j \rVert_{\mathbb{L}^2}^2 
    \right)  \Bigg] \ .
\end{split}
\end{align} 
Note, the first loss of the minimization problem is the Euclidean-vector norm and not the functional $\mathbb{L}^2$ norm. 

\subsection{Matrix representation}
\label{subsec:matrix_scala_on_function}

One can use the FPCs of each feature $\mathcal X_j$ in order to obtain a matrix representation of \eqref{eq:functional_minimization_problem_scalar}.
Following the steps described in Section \ref{subsec:coeff_representation}, let us consider the simple case with just one feature $\mathcal{X}_1$. Let $\Xi_1 = \left[\Xi_{11} ~ | ~ \dots ~ | ~ \Xi_{1L} \right]$ be the matrix containing the first $L$ FPCs of $\mathcal{X}_1$.
Let $X_{[1]} \in \mathbb{R}^{n \times L}$ be the score matrix of $\mathcal X_1$ with respect to $\Xi_1$, and $B_{1} \in \mathbb{R}^{L}$ be the score vector of $\mathcal B_1$ with respect to the same basis system $\Xi_1$. 
Then, we can approximate $\mathcal X_1(s) \approx X_{[1]}\Xi_1(s)^T, ~ \mathcal B_1(s) \approx \Xi_1(s)B_{1}$, and \eqref{eq:scalar_model} as
 \begin{align*}
 \small
 \begin{split}
     Y \approx X_{[1]} \left(\int_{\mathcal{S}} \Xi_1(s)^T\Xi_1(s)\,ds\right) B_{1} \ ,
 \end{split}
 \end{align*}
Recalling that $\int \Xi_1(s)^T\Xi_1(s)\,ds = I_L$, this gives us
$Y \approx X_{[1]} B_{1}$. The extension to $p$ predictors is straightforward. For each predictor $\mathcal{X}_j$, $j=1,\dots,p$, define $X_{[j]}$ as the score matrix of $\mathcal{X}_{j}$ with respect to $\Xi_j = \left[\Xi_{j1} ~ | ~ \dots ~ | ~ \Xi_{jL} \right]$. Similarly, denote the score vector of $\mathcal B_j$ with respect to the same basis system $\Xi_j$ as $B_{j} \in \mathbb{R}^{L}$.
Let us indicate the design matrix as $X = \left[X_{[1]} | \dots | X_{[p]} \right] \in \mathbb{R}^{n \times pL}$,
the coefficient vector as $B = \left(B_{1}^T, \dots , B_{p}^T \right)^T \in \mathbb{R}^{pL}$. Thus, model \eqref{eq:scalar_model} can be approximated as $ Y \approx X B $, and the matrix representation of the minimization problem \eqref{eq:functional_minimization_problem_scalar} is:
 \begin{equation}
 \small
     \label{eq:matrix_minimization_problem_scalar}
     \min_{B} \frac{1}{2}\lVert Y-XB\rVert_{2}^2 + \sum_{j=1}^{p} \omega_j \left( \lambda_1 \lVert B_{j}\rVert_{2}+\frac{\lambda_2}{2} \lVert B_{j}\rVert_{2}^2 \right) \ .
 \end{equation}
The only differences with respect to the matrix representation of the function-on-function scenario \eqref{eq:matrix_minimization_problem} are the dimensions of $Y$ and $B$. Once an estimate $\check B$ of the score vector is obtained, one can recover the coefficient curves as
\begin{equation*}
\small
\label{eq:from_B_to_funB_scalar}
\hat{\mathcal B}_j(s) = \Xi_j(s)\check B_j(s) \ \ \ j=1,\dots,p \ .
\end{equation*}

\subsection{FAStEN implementation}

Here we describe how to implement the \texttt{FAStEN} algorithm for the scalar-on-function minimization problem.

\subsubsection*{Primal and Dual formulation}
The primal and dual formulation of the minimization problem \eqref{eq:matrix_minimization_problem_scalar} are 
\begin{align*}
 \small
 \begin{split}
       \min_B h(XB) + \pi(B) \ \ \ \ \text{and} \ \ \ \ \min_{V,Z} h^*(V) + \pi^*(Z) \ \ s.t. \ \ X^TV+Z=0 \ ,
 \end{split}
 \end{align*}
respectively. $h(XB)= \frac{1}{2}\lVert Y-XB\rVert_{2}^2$, $ \ \pi(B) = \sum_{j=1}^{p} \omega_j 
\left( \lambda_1 \lVert B_{j}\rVert_{2} +\frac{\lambda_2}{2} \lVert B_{j}\rVert_{2}^2 \right)$. $ \ V\in\mathbb{R}^{n}$ and $Z\in\mathbb{R}^{pL}$ are the vector dual variables. Using the same notation introduced for $B$, we can write $Z = \left(Z_1^T, \dots , Z_{p}^T \right)^T$, where $Z_{j}\in \mathbb{R}^{L}$ is the sub-vector of $Z$ associated with the $j$-th feature. The conjugate function $h^*$ has the form $h^*(V) = \frac{1}{2}\lVert V\rVert_2^2 + Y^T V$, while the from of $\pi^*$ is given in the next paragraph. \\

\noindent
\emph{Conjugate function and proximal operator.} 
Both the conjugate function $\pi^*$ and the proximal operator $\prox_{\sa \pi}$ depends on the sub-vectors $Z_j$ and $B_j$, respectively. Thus, we can use the results in \cite{boschi2021highly}, where these mathematical operators have been computed for vectors in the case of the function-on-scalar regression model. Specifically, we can incorporate the weights $\omega$'s in the penalty parameters $\la_1$ and $\la_2$, and obtain: 
\begin{align*}
 \small
 \begin{split}
       \pi^*(Z) = \sum_{j=1}^p \pi^*(Z_j) =  \sum_{i=1}^p  (2 \omega_j \la_2)^{-1} \left(\big[\norm{Z_j}_2 - \omega_j\la_1\big]_{+}\right)^2 \ ,
 \end{split}
 \end{align*}
and $\prox_{\sigma \pi}(B) = \big(\prox_{\sigma \pi}(B_1)^T, \dots, \prox_{\sigma \pi}(B_p)^T \big)^T$, where     
\begin{align*}
 \small
 \begin{split}
      \prox_{\sigma \pi}(B_j) =  (1 + \sa \omega_j \la_2)^{-1} \left[1- \norm{B_j}_2^{-1} \sa \omega_j \la_1  \right]_{+} B_j.
 \end{split}
 \end{align*}
Note, $\pi^* : \mathbb{R}^{pL} \rightarrow \mathbb{R}$ and  $\prox_{\sigma \pi}(B) : \mathbb{R}^{pL} \rightarrow\mathbb{R}^{pL}$. To compute $\prox_{\pi^*/\sa}(B)$, one can use the {Moreau decomposition}:
$\prox_{\pi^*/\sa} \left(B\right) = B/\sa - \prox_{\sa p}\left(B\right)/\sa$. \\

\noindent
\emph{Dual Augmented Lagrangian and KKT.} The Dual Augmented Lagrangian is 
\begin{align*}
\small
\begin{split}
    \mathcal{L}_\sigma (V,Z,B) =  h^*(V) + \pi^*(Z) - 
     \sum_{j=1}^p \langle B_{j}, X_{[j]}^T V + Z_{j} \rangle + 
    \frac{\sigma}{2} \sum_{j=1}^p \lVert X_{[j]}^T V + Z_{j}\rVert_2^2
\end{split}
\end{align*}
The KKT equations have the same form of \eqref{eq:kkt} but the arguments have different dimensions. 

\subsubsection*{Z and V update} 
\emph{Z update}. Following the same steps of Proposition \ref{prop:z_update} proof, it is possible to show that: 
\begin{align*}
\small
\begin{split}
    \Bar{Z} = \arg\min_Z \mathcal{L}_\sigma (Z\,|\,\Bar{V},B) = \prox_{\pi^*/\sa} \left(B/\sa - X^T\Bar{V} \right).
\end{split}
\end{align*} 

\noindent
\emph{V update}. To update V, we perform the Newton method: $V^{m+1} = V^m + u D$, where $u$ is the step size and  $D \in \mathbb{R}^{n \times 1}$ the descent direction computed by solving the linear system
\begin{equation*}
\small
    H_{\psi}(V) D = -\nabla \psi (V) \ .
\end{equation*}
$H_{\psi} \in \mathbb R^{n \times n}$ is the Hessian matrix, 
$\nabla \psi \in \mathbb R^{n}$ is the gradient vector. Note that, differently from function-on-function case, the dimension of the Hessian and the gradient does not depend on $L$. The following result is the equivalent of Theorem \ref{th:v_update}. The proof follows the same steps, but in this case we are dealing with lower order tensors and operators. 
\begin{proposition}
Let $T=B-\sa X^T V$, $T_{j}=B_{j} - X_{[j]}^T V$, $\mathcal{J} = \big\{ j~:~ \lVert T_{j} \rVert_2  \ge \sa \omega_j \la_1 \big\}$, and $r = |\mathcal{J}|$ be the cardinality of $\mathcal{J}$.
Next, let $ X_\mathcal{J} \in \mathbb{R}^{n \times rL}$ be the sub-matrix of $X$ restricted to the blocks $X_{j}$, 
$j \in \mathcal{J}$.
Define the squared $L \times L$ matrix 
    \begin{equation*}
    \small
        P_{[j]} = (1 + \sa \omega_j \la_2)^{-1} \left( \left(1 -
        \frac{\sa \omega_j \la_1}{ \norm{T_{j}}_2} \right) I_{L^2} +
        \frac{\sa \omega_j \la_1}{ \norm{T_{j}}_2^3} 
        T_{j}T_{j}^T \right) \ .
    \end{equation*}
Finally, let $Q_\mathcal{J} \in \mathbb{R}^{rL \times rL}$ be the block-diagonal matrix formed by the blocks $P_{[j]}$, $j \in \mathcal{J}$.
Then
  \begin{align*}
    \small
    \begin{split}
        &(i)~ \  \ \psi(V) = h^*(V) + \frac{1}{2\sa}  \sum_{j=1}^p 
        \left( \left( 1 + \sa \omega_j \la_2 \right) \normbig{\prox_{\sa \pi}\big(T_{j}\big) }_2^2 - \norm{B_{j}}_2^2 \right) \\ 
        &(ii)~ \ \nabla \psi(V) = V + Y - X \prox_{\sa \pi}(T) \\
        &(iii)~H_{\psi}(V) = I_{n} + \sa X_\mathcal{J} Q_\mathcal{J} X_\mathcal{J}^T \ .
    \end{split}
    \end{align*}
\end{proposition}
\vspace{0.4cm}
\noindent
\emph{Computational cost.} Selecting a subset of active $r$ features at each iteration reduces the total cost of solving the linear system from $\mathcal{O} \l( n(n^2 + npL + p^2L^2) \r)$ to $\mathcal{O} \l( n(n^2 + nrL + r^2L^2) \r)$.
When $r$ is smaller than $n$, one can perform the inversion through the \emph{Sherman-Morrison-Woodbury} formula:
\begin{align*}
    \small
    \begin{split}
       \left(I_{n} + \sa X_\mathcal{J} Q_\mathcal{J} X_\mathcal{J}^T \right)^{-1} = 
       I_{n} -  X_\mathcal{J}  \left( \l( \sa Q_\mathcal{J} \r)^{-1} + X_\mathcal{J}^T X_\mathcal{J} \right)^{-1} X_\mathcal{J}^T
    \end{split}
\end{align*}
which allows one to factorize an $rL \times rL$ matrix. The total cost after some manipulation is  $\mathcal{O} \l( rL(L^2 + nrL + r^2L^2 + n^2) \r)$.

{\green 
\section{Algorithmic convergence analysis}
\label{sec:supp_conv}

In this section, we establish the convergence rate of FAStEN for both the Dual Augmented Lagrangian and the inner sub-problem. The proof adheres to the steps outlined in \cite{tomioka2009dual, li2018, boschi2020efficient}.

\subsection{Dual Augmented Lagrangian}

In order to establish the global convergence of \textbf{Algorithm \ref{alg:dal}} and the super-linear convergence of the solution, we refer to \emph{Theorem 3.2} and \emph{Theorem 3.3} in \citet{li2018},  which are founded on the crucial results presented in \citet{rockafellar1976augmented, rockafellar1976, luque1984asymptotic}.
We proceed by verifying that the theorem's assumptions hold. In particular, we met the assumptions on $h(\cdot)$. Moreover, it is always feasible to implement the stopping criteria for the local convergence analysis delineated in \citet{li2018} -- Section 3. We have to verify that the operators $\mathcal T_f$ and $\mathcal T_l$ fulfill the \emph{error bound condition}.

\noindent
Let $\mathcal{F}$ represent the objective function \eqref{eq:matrix_minimization_problem}. Note that $\mathcal{F}$ is a closed proper convex function, while the Dual Augmented Lagrangian $\mathcal{L}$ function \eqref{eq:augmented_lagrangian} is convex-concave. We define the maximal monotone operators $\mathcal T_\mathcal{F}$ and $\mathcal T_\mathcal{L}$ as in \citet{rockafellar1976augmented}:
\begin{equation}
\small
\label{eq:t_operators}
    \mathcal{T}_\mathcal{F}(B) = \nabla \mathcal{F}(B), \quad 
    \mathcal{T}_\mathcal{L} (V, Z, B) = \{ (V', Z', B') | (V', Z', -B') \in \partial \mathcal{L}(V, Z, B)\}.
\end{equation}
We need to prove that $\mathcal T_\mathcal{F}$ and $\mathcal T_\mathcal{L}$ are \emph{metric subregular} \citep{dontchev2009implicit}, which is equivalent to satisfying the error bound condition the \emph{error bound condition} \citep{robinson1981some}, or the \emph{growth condition} \citep{luque1984asymptotic}. In particular, we say that a multivalue mapping $G~:~ \text{dom}(\mathcal{F}) \rightrightarrows \text{dom}(\mathcal{L})$ satisfies the \emph{error bound condition} at $B \in \text{dom}(\mathcal{F})$ with modulus $\kappa > 0$ if $G^{-1}(V) \ne \emptyset$ and there exists $\eta > 0$ such that if $B \in \text{dom}(B)$ with $\text{dist}(V, G(B)) \le \eta$, then
\begin{equation}
\small
\label{eq:error_bound_cond}
    \text{dist}(B, G^{-1}(V)) \le \eta \text{dist}(V, G(B)).
\end{equation}
The regularity of $\mathcal{T}_\mathcal{F}$ can be deduced by \citet{zhou2017unified}: as $\nabla h$ is  Lipschitz continuous and $\pi$ possesses a polyhedral epigraph, $\mathcal{T}_\mathcal{F}$ complies with the error bound condition. 
In order to confirm the bound condition for  $\mathcal{T}_\mathcal{L}$,  we can employ several established results, given the special form of $\pi^*$ in \eqref{eq:p_star}. It is important to note that $\pi^*$ is a piecewise linear-quadratic function. Thus, we can apply \emph{Proposition 12.30} in \citet{rockafellar2009variational} and assert that the gradient mapping $\nabla \pi^*$ is piecewise polyhedral. 
Finally, polyhedral multi-functions satisfy the error bound condition for any point $V \in \text{dom}(\mathcal{L})$ \citep{robinson1981some}. This proves the regularity of $\mathcal{T}_\mathcal{L}$ and, consequently, the super-linear convergence of the method.

\subsection{Inner sub-problem}

To state the super-linear convergence of the sequence $\{V^m\}$ generated by the \emph{Newton Method} in \textbf{Algorithm \ref{alg:dal}}, we refer to \emph{Theorem 3.6} in \citet{li2018}, which is grounded on the pivotal findings in \citet{zhao2010newton}. Verifying all the assumptions is straightforward. $\nabla h^*$ and $\prox_{\sa \pi}$ in \eqref{eq:prox_op} are smooth functions. 
According to \emph{Proposition 3.3} in \citet{zhao2010newton}, $D$ defined in \eqref{eq:newton_direction} is a descent direction. Lastly, by recalling the definition of $H_{\psi}$ in Theorem \ref{th:v_update}, we can observe that it is positive semidefinite.
}

%
%
{\green 
\section{Asymptotic results}
\label{sec:supp_asympt}

To prove Theorem \ref{th:oracle}, we first state some useful Lemmas. We follow \cite{kong2016partially, cai2021variable, zhang2022subgroup} in the remainder of this Section, adapting their results to our framework and extending them to the high-dimensional function-on-function regression setting. 

\noindent
Lemma \ref{lemma:FPCrates} provides asymptotic rates for FPC scores and functions. It states that the differences between true and estimated FPCs are asymptotically negligible.

\begin{lemma}
	\label{lemma:FPCrates}
    Under conditions \textbf{A}, \textbf{B}, and \textbf{D}, for $j,j_1,j_2 = 1,\dots,p$, $\ell,\ell_1,\ell_2 = 1,\dots,L$, $k = 1,\dots,K$, and $i=1,\dots,n$, we have:
    \begin{enumerate}[label=(\roman*)]
        \item $|{X}_{[j](i\ell)} - \mathring{X}_{[j](i\ell)}|\xi_{j\ell}^{-1} = O_p(n^{-1} \ell^{\alpha_2 + 2})$ and $|{Y}_{ik} - \mathring{Y}_{ik}| = O_p(n^{-1} k^{2})$\,;
        
        \item $\left|\sum_{i=1}^n ({X}_{[j_1](i\ell_1)} {X}_{[j_2](i\ell_2)} - \mathring{X}_{[j_1](i\ell_1)} \mathring{X}_{[j_2](i\ell_2)})(\xi_{j_1\ell_1} \xi_{j_2\ell_2})^{-1/2} \right| = O_p(\ell_1^{\alpha_2/2 + 1} n^{1/2} + \ell_2^{\alpha_2/2 + 1} n^{1/2})$\,;
        
        \item $\left|\sum_{i=1}^n (\mathring{X}_{[j_1](i\ell_1)} \mathring{X}_{[j_2](i\ell_2)} - \EE[\mathring{X}_{[j_1](i\ell_1)} \mathring{X}_{[j_2](i\ell_2)}])(\xi_{j_1\ell_1} \xi_{j_2\ell_2})^{-1/2} \right| = O_p(n^{1/2})$\,;

        \item $\left|\sum_{i=1}^n ({X}_{[j_1](i\ell_1)} {X}_{[j_2](i\ell_2)} - \EE[\mathring{X}_{[j_1](i\ell_1)} \mathring{X}_{[j_2](\ell_2)}])(\xi_{j_1\ell_1} \xi_{j_2\ell_2})^{-1/2} \right| = O_p(\ell_1^{\alpha_2/2 + 1} n^{1/2} + \ell_2^{\alpha_2/2 + 1} n^{1/2})$\,.
    \end{enumerate}

\end{lemma}

\noindent\noindent \textbf{\textit{Proof}}
See Lemmas 1 and 2 in \cite{kong2016partially} or Lemma 1 in \cite{cai2021variable, zhang2022subgroup} for a proof. Results follow from standard theory on FPCs \citep{horvath2012inference}.

\noindent
Define $\Tilde{Z}$ as the $n\times p_0L$ matrix whose $i$-th row is $\Tilde{Z}_i$. Similarly, define $\hat{Z} $ as the empirical counterpart of $\Tilde{Z}$, where FPC scores are replaced by their estimated values. Moreover, define the $nK\times p_0KL$ matrices $\Tilde{N} = \Tilde{Z} \otimes I_K$ and $\hat{N} = \hat{Z} \otimes I_K$. Lemma \ref{lemma:eigenvalues} shows that the minimum and maximum eigenvalues of $\hat{N}^T\hat{N}/n$ are consistent with the ones of $\EE[(\Tilde{Z}_i \otimes I_K)(\Tilde{Z}_i \otimes I_K)^T]$.

\begin{lemma}
    \label{lemma:eigenvalues}
    Under conditions \textbf{A}, \textbf{B}, \textbf{E}, \textbf{F}, and \textbf{G} we have:
     \begin{enumerate}[label=(\roman*)]
        \item $\left|\rho_{min}\left(\frac{\hat{N}^T\hat{N}}{n}\right) - \rho_{min}(\EE[(\Tilde{Z}_i \otimes I_K)(\Tilde{Z}_i \otimes I_K)^T]) \right| = o_p(1)$;
     	
        \item $\left|\rho_{max}\left(\frac{\hat{N}^T\hat{N}}{n}\right) - \rho_{max}(\EE[(\Tilde{Z}_i \otimes I_K)(\Tilde{Z}_i \otimes I_K)^T]) \right| = o_p(1)$.
    \end{enumerate}
\end{lemma}

 \noindent \textbf{\textit{Proof}}
    We first show $(i)$. By Lemma \ref{lemma:FPCrates}, we know that:
    \begin{equation*}
    \small
        \begin{split}
            \left|\rho_{min}\left(\frac{\hat{Z}^T\hat{Z}}{n}\right) - \rho_{min}(\EE[\Tilde{Z}_i\Tilde{Z}_i ^T]) \right| &\leq \left\lVert \frac{\hat{Z}^T\hat{Z}}{n} - \EE[\Tilde{Z}_i \Tilde{Z}_i^T] \right\rVert_1 \\
            &\leq O_p\left( p_0 \sum_{\ell=1}^L n^{-1/2}\ell^{\alpha_2/2 + 1} + n^{-1/2}L^{\alpha_2/2 + 1} \right) \\
            &\leq O_p(p_0 L^{\alpha_2/2 + 2} n^{-1/2})\,.
        \end{split}
    \end{equation*}
    Now, notice that $\hat{N}^T\hat{N} = (\hat{Z} \otimes I_K)^T(\hat{Z} \otimes I_K) = (\hat{Z}^T\hat{Z}) \otimes I_K$ and $\EE[(\Tilde{Z}_i \otimes I_K)(\Tilde{Z}_i \otimes I_K)^T] = \EE[\Tilde{Z}_i\Tilde{Z}_i^T] \otimes I_K$. Thus, $\rho_{min}(\hat{N}^T\hat{N}/n) = \rho_{min}(\hat{Z}^T\hat{Z}/n)$ and $\rho_{min}(\EE[(\Tilde{Z}_i \otimes I_K)(\Tilde{Z}_i \otimes I_K)^T]) = \rho_{min}(\EE[\Tilde{Z}_i\Tilde{Z}_i^T])$. 
    Therefore, by condition \textbf{D}, we conclude that
    \begin{equation*}
        \small
        \left|\rho_{min}\left(\frac{\hat{N}^T\hat{N}}{n}\right) - \rho_{min}(\EE[(\Tilde{Z}_i \otimes I_K)(\Tilde{Z}_i \otimes I_K)^T]) \right| = O_p(p_0 L^{\alpha_2/2 + 2}n^{-1/2}) = o_p(1)\,.
    \end{equation*}
    \noindent
    The proof of $(ii)$ is analogous to the one of $(i)$, \textit{mutatis mutandis}.

\noindent
Define the $nK \times nK$ projection matrix $P = \hat{N}(\hat{N}^T \hat{N})^{-1} \hat{N}^T$. Also, denote the vectorized true and estimated FPCs scores of the response variable as $\mathring{\Lambda}$ and ${\Lambda}$ (these are $nK\times1$ vectors). Moreover, define $\Tilde{b}_0$, the $p_0KL\times1$ vector of standardized coefficients $\Tilde{B}_{[j](k\ell)}$ associated with the set of active features. Finally, define the $nK \times 1$ vectors $\mathring{\Delta} = P(\mathring{\Lambda} - \hat{N}\Tilde{b}_0)$ and ${\Delta} = P({\Lambda} - \hat{N}\Tilde{b}_0)$.
Lemma \ref{lemma:projection_rates} concerns the asymptotic order of $\hat{\Delta}$. 
\begin{lemma}
	\label{lemma:projection_rates}
    Let $\tau^2 = K^3 + p_0(KL + nL^{-2\gamma_2 - \alpha_2})$. Under conditions \textbf{A}, \textbf{B}, \textbf{C}, \textbf{D}, \textbf{F}, and \textbf{G}, we have:
    \begin{equation*}
    	\small
    	\lVert {\Delta} \rVert_2^2 = O_p(\tau^2)\,.
    \end{equation*}
\end{lemma}
 \noindent \textbf{\textit{Proof}}
	By Condition \textbf{F} and Lemma \ref{lemma:eigenvalues}, we know that $\hat{N}^T\hat{N}$ is invertible and thus $P$ exists. We first explore the asymptotic behavior of $\mathring{\Delta}$:
	\begin{equation*}
		\begin{split}
				\mathring{\Delta} = P(\mathring{\Lambda} - \hat{N}\Tilde{b}_0) &= P(\varepsilon + \nu + \Tilde{N}\Tilde{b}_0 - \hat{N}\Tilde{b}_0) \\
				&= P\big(\varepsilon + \nu + (\Tilde{N} - \hat{N})\Tilde{b}_0\big)\,,
		\end{split}
	\end{equation*}
	where $\varepsilon = [\mathring{\varepsilon}_{11}, \dots, \mathring{\varepsilon}_{nK}]^T$ is the $nK\times1$ vector containing the FPC scores of the error term w.r.t.~the true FPCs of the response variable; $\nu = [\nu_{11},\dots,\nu_{nK}]^T$ is the $nK\times1$ vector whose generic element is $\nu_{ik} = \sum_{j=1}^p \sum_{\ell=L+1}^\infty \mathring{X}_{[j](i\ell)} \mathring{B}_{[j](k\ell)} = \sum_{j=1}^{p_0} \sum_{\ell=L+1}^\infty  \mathring{X}_{[j](i\ell)} \mathring{B}_{[j](k\ell)} $. 
\noindent
We now analyze each term independently. We start with $P\varepsilon$. Notice that $\EE[\lVert P\varepsilon\rVert_2^2] = \EE[\varepsilon^TP\varepsilon] = \EE[tr(P\EE[\varepsilon\varepsilon^T])]$. By condition \textbf{A} and orthonormality of $\{\mathring{\Upsilon}_k:\,k\in\NN \}$, it follows that $\EE[\mathring{\varepsilon}_{ik}^2] \leq \EE[\cnorm{\epsilon_i(t)}^2] \leq C$ and $\EE[\mathring{\varepsilon}_{i_1k_1}\mathring{\varepsilon}_{i_2k_2}] = 0$ for any $i_1\neq i_2$ or $k_1\neq k_2$. Thus, $\EE[\lVert P\varepsilon\rVert_2^2] \leq C tr(P) = O(p_0LK)$. Therefore, $\lVert P\varepsilon \rVert_2^2 = O_p(p_0KL)$. 
	
\noindent 
We move to $P\nu$. By exploiting conditions \textbf{C} and \textbf{B}, we get:
	\begin{equation*}
		\small
		\begin{split}
			\EE[\lVert P\nu\rVert_2^2] \leq \EE[\lVert \nu \rVert_2^2] &\leq \EE\left[\sum_{i=1}^n \sum_{k=1}^K \left( \sum_{j=1}^{p_0} \sum_{\ell=L+1}^\infty \mathring{X}_{[j](i\ell)} \mathring{B}_{[j](k\ell)} \right)^2 \right] \\
			&\leq O\left(\sum_{i=1}^n \sum_{k=1}^K  \sum_{j=1}^{p_0} \sum_{\ell=L+1}^\infty \EE \left[ \mathring{X}_{[j](i\ell)}^2 \mathring{B}_{[j](k\ell)}^2 \right] \right) \\
			&\leq O\left(\sum_{i=1}^n \sum_{k=1}^K  \sum_{j=1}^{p_0} \sum_{\ell=L+1}^\infty \EE [ \mathring{X}_{[j](i\ell)}^2] k^{-2\gamma_1} \ell^{-2\gamma_2} \right) \\
			&\leq O\left(\sum_{i=1}^n \sum_{k=1}^K  \sum_{j=1}^{p_0} \sum_{\ell=L+1}^\infty \xi_{j\ell} k^{-2\gamma_1} \ell^{-2\gamma_2} \right) \\
			&\leq O\left(\sum_{i=1}^n \sum_{k=1}^K  \sum_{j=1}^{p_0} \sum_{\ell=L+1}^\infty  k^{-2\gamma_1} \ell^{-2\gamma_2 - \alpha_2} \right) \\
			&\leq O\left(np_0  \sum_{k=1}^K   \sum_{\ell=L+1}^\infty k^{-2\gamma_1} \ell^{-2\gamma_2 - \alpha_2} \right)\,,
		\end{split}
	\end{equation*}
	Thus, $\lVert P\nu \rVert_2^2 = O_p(np_0 L^{-2\gamma_2 - \alpha_2})$. 
	
	Finally, we focus on $P(\Tilde{N} - \hat{N})\Tilde{b}_0$. By Lemma \ref{lemma:FPCrates} and condition \textbf{C}, we get:
	\begin{equation*}
		\small
		\begin{split}
			\lVert P(\Tilde{N} - \hat{N})\Tilde{b}_0\rVert_2^2 \leq \lVert (\Tilde{N} - \hat{N})\Tilde{b}_0 \rVert_2^2 &\leq \sum_{i=1}^n \sum_{k=1}^K \left[ \sum_{j=1}^{p_0} \sum_{\ell=1}^L (\mathring{X}_{[j](i\ell)} - X_{[j](i\ell)}) \mathring{B}_{[j](kl)} \right]^2 \\
			&\leq O \left(\sum_{i=1}^n \sum_{k=1}^K  \sum_{j=1}^{p_0} \sum_{\ell=1}^L (\mathring{X}_{[j](i\ell)} - X_{[j](i\ell)})^2 \mathring{B}_{[j](kl)}^2\right) \\
			&\leq O \left(\sum_{i=1}^n \sum_{k=1}^K  \sum_{j=1}^{p_0} \sum_{\ell=1}^L (\mathring{X}_{[j](i\ell)} - X_{[j](i\ell)})^2 k^{-2\gamma_1} \ell^{-2\gamma_2}\right) \\
			&\leq O \left(\sum_{i=1}^n \sum_{k=1}^K  \sum_{j=1}^{p_0} \sum_{\ell=1}^L n^{-1} k^{-2\gamma_1} \ell^{2-2\gamma_2}\right) \\
			&\leq O \left(p_0 \sum_{k=1}^K  \sum_{\ell=1}^L  k^{-2\gamma_1} \ell^{2-2\gamma_2}\right)\,.
		\end{split}
	\end{equation*}
	Thus, $P(\Tilde{N} - \hat{N})\Tilde{b}_0 = O_p(p_0)$. \\
	By joining the previous results, we get $\lVert \mathring{\Delta} \rVert_2^2 = O_p(p_0KL + np_0L^{-2\gamma_2 - \alpha_2})$. \\
	By Lemma \ref{lemma:FPCrates}, we get that $\lVert \mathring{\Delta} - {\Delta}\rVert_2^2 \leq \lVert \mathring{\Lambda} - {\Lambda} \rVert_2^2 = O_p(K^3)$. Thus, $\lVert \Delta \rVert_2^2 \leq \lVert \mathring{\Delta} - \Delta \rVert_2^2 + \lVert \mathring{\Delta} \rVert_2^2 = O_p(K^3 + p_0KL + np_0L^{-2\gamma_2 - \alpha_2}) = O_p(\tau^2)$.

\noindent
A direct implication of the previous Lemma is contained in the following Corollary, which concerns the asymptotic order of the $n\times1$ subvectors $\Delta_k$ of $\Delta$. 
\begin{corollary}
    \label{corollary:reduced_projection_rates}
	 Let $\tau_k^2 = K^2 + p_0(L + nL^{-2\gamma_2 - \alpha_2})$. Under conditions \textbf{A}, \textbf{B}, \textbf{C}, \textbf{D}, \textbf{F}, and \textbf{G}, we have:
	\begin{equation*}
		\small
		\lVert {\Delta}_k \rVert_2^2 = O_p(\tau_k^2)\,.
	\end{equation*}
\end{corollary}

\subsection{Theorem \ref{th:oracle}}
We are now ready to prove Theorem \ref{th:oracle}. The idea of the proof is the following: consider the oracle estimator $(\hat{B}_0, 0)^T$ as a candidate solution of the objective function in (\ref{eq:matrix_minimization_problem}). It clearly satisfies $\PP[\hat{B}_{j} = 0,\, j=p_0+1,\dots,p] \to 1$. First, we show that the $p_0KL$ elements of $\hat{B}_0$ associated with the active features are a minimizer of the constrained problem in the $p_0$ dimensional space, and satisfy $\lVert \hat{B}_0 - \Tilde{B}_0 \rVert_2^2 = O_p(r^2/n)$. Then, we show that the oracle estimator is a local minimizer over the whole $p$ dimensional space. Denote the objective function in (\ref{eq:matrix_minimization_problem}) as $\mathcal{L}$, and its constrained version on the $p_0$ dimensional space as $\mathcal{L}_0$. \\

 \noindent \textbf{\textit{Proof}}
The proof proceeds in two steps. First, we show that the oracle estimator $\hat{B}_0$ is consistent in estimation over the reduced space spanned by the set of $p_0$ active features. Second, we show that the oracle estimator is also a local minimizer of (\ref{eq:matrix_minimization_problem}) over the whole $p$ dimensional space.

\noindent
In order to show the first point, it is enough to prove that, for any $\delta$, there exists $C$ such that $\PP[\inf_{\lVert u \rVert_2 = C} \mathcal{L}_0(\Tilde{B}_0 + \alpha u) > \mathcal{L}_0(\Tilde{B}_0)] > 1 - \delta$. By Lemma \ref{lemma:eigenvalues}, we have:
\begin{equation*}
    \small
    \begin{split}
        \mathcal{L}_0(\Tilde{B}_0 + \alpha u) - \mathcal{L}_0(\Tilde{B}_0) &= \lVert \Lambda - \hat{N}(\Tilde{B}_0 + \alpha u) \rVert_2^2 - \lVert \Lambda - \hat{N} \Tilde{B}_0 \rVert_2^2 \\
        &\geq \lVert \alpha\hat{N}u \rVert_2^2 - 2\alpha\Delta^T\hat{N}u \\ 
        &\geq n\alpha^2\rho_{min}\left(\frac{\hat{N}^T\hat{N}}{n}\right) \lVert u \rVert_2^2 - 2\alpha\lVert \Delta \rVert_2 \rho_{max}^{1/2}\left(\frac{\hat{N}^T\hat{N}}{n}\right) \lVert u \rVert_2 n^{1/2} \\ 
        &\geq n \alpha^2 C_1 \lVert u \rVert_2^2 - n^{1/2} \alpha C_2 \lVert\Delta\rVert_2 \lVert u \rVert_2\,.
    \end{split}
\end{equation*}
The second term of the last inequality, by Lemma \ref{lemma:projection_rates}, is $O_p(n\alpha^2\lVert u\rVert_2)$; thus, allowing $\lVert u\rVert_2 = C$ to be large enough, the first term dominates the second. We have proved that there exists a local minimizer $\hat{B}_0$ of $\mathcal{L}_0$ such that $\lVert \hat{B}_0 - \Tilde{B}_0 \rVert_2^2 = O_p(r^2/n)$.

\noindent
We now need to show that the oracle estimator $(\hat{B}_0, 0)^T$ is also a local minimizer of $\mathcal{L}$, the objective function over the whole $p$ dimensional space. By the Karush-Kuhn-Tucker conditions \citep[Sec.~5.5.3]{boyd2004convex}, it is sufficient to show that $(\hat{B}_0, 0)^T$ satisfies
     \begin{enumerate}[label=(\roman*)]
        \item $\min_{j\in\{1,\dots,p_0\}} \lVert B_j \rVert_2 \geq C \lambda_1 \omega_j$, for $j\in\{1,\dots,p_0\}$;
     	
        \item $\lVert S_j(\hat{B}) \rVert_2 \leq \lambda_1 \omega_j n$, for $j\notin\{1,\dots,p_0\}$;
    \end{enumerate}
    where 
    \begin{equation*}
        \small
        S_j(\hat{B}) = \frac{\partial \sum_{i=1}^n \sum_{k=1}^K  \left(Y_{ik} - \sum_{j=1}^p \sum_{\ell=1}^{L} X_{[j](i\ell)} B_{[j](k\ell)} \right)^2}{\partial \hat{B}_{[j]}}\,.
    \end{equation*}

\noindent
As for (i), notice that $\lVert B_{[j]} \rVert_2 \geq \lVert \mathring{B}_{[j]} \rVert_2 - \lVert B_{[j]} - \mathring{B}_{[j]} \rVert_2 $ and, therefore, by condition \textbf{E}, $\min_{j\in\{1,\dots,p_0\}} \lVert \mathring{B}_{[j]} \rVert_2 / (\lambda_1 \omega_j) \to \infty$. Since $\lVert \hat{B}_0 - \Tilde{B}_0 \rVert_2 = O_p(r/n^{1/2})$ and $\Tilde{B}_{[j](k\ell)} = \xi_{j\ell}^{1/2} \mathring{B}_{[j](k\ell)}$, we have $\lVert {B}_0 - \mathring{B}_0 \rVert_2 = O_p(rL^{\alpha_2/2}/n^{1/2}) = o_p(\lambda_1\omega_j)$. Thus $\min_{j\in\{1,\dots,p_0\}} \lVert {B}_{[j]} \rVert_2 / (\lambda_1 \omega_j) \to \infty$ in probability, which implies $\PP[\lVert B_{[j]} \rVert_2 \geq C \lambda_1 \omega_j,\, j\in\{1,\dots,p_0\}]\to1$.

\noindent
We now move to (ii). We need to show that $\PP[\max_{j\notin\{1,\dots,p_0\}} \lVert S_j(\hat{B}) \rVert_2 > \lambda_1 \omega_j n] \to 0$. Denote the $(k,\ell)$-th element of $S_j(\hat{B})$ as $S_{[j](k\ell)}(\hat{B})$. We have:
\begin{equation*}
    \small
    \begin{split}
        S_{[j](k\ell)}(\hat{B}) &= \frac{\partial \sum_{i=1}^n \sum_{k=1}^K  \left(Y_{ik} - \sum_{j=1}^p \sum_{\ell=1}^{L} X_{[j](i\ell)} B_{[j](k\ell)} \right)^2}{\partial \hat{B}_{[j](k\ell)}} \\
        &= \frac{\partial \sum_{i=1}^n \sum_{k=1}^K  \left(Y_{ik} - \sum_{j=1}^{p_0} \sum_{\ell=1}^{L} X_{[j](i\ell)} B_{[j](k\ell)} \right)^2}{\partial \hat{B}_{[j](k\ell)}} \\
        &= -2 \sum_{i=1}^n X_{[j](i\ell)} \left(Y_{ik} - \sum_{j=1}^{p_0} \sum_{\ell=1}^L X_{[j](i\ell)} \hat{B}_{[j](k\ell)} \xi_{j\ell}^{-1/2} \right) \\
        &= -2 X_{[j](\cdot\ell)}^T \Delta_k\,,
    \end{split}
\end{equation*}
where $X_{[j](\cdot\ell)}$ is a $n\times1$ vector whose $i$-th element is $X_{[j](i\ell)}$. Thus we can write:
\begin{equation*}
    \small
    \begin{split}
        \PP\left[\max_{j\notin\{1,\dots,p_0\}} \lVert S_j(\hat{B}) \rVert_2 > \lambda_1 \omega_j n \right] &= \PP\left[\max_{j\notin\{1,\dots,p_0\}} \bigg(\sum_{k=1}^K \sum_{\ell=1}^L S_{[j](k\ell)}^2(\hat{B})\bigg)^{1/2} > \lambda_1 \omega_j n \right] \\
        &\leq \PP\left[\max_{j\notin\{1,\dots,p_0\}} \max_{k\in\{1,\dots,K\}} \max_{\ell\in\{1,\dots,L\}} \lvert S_{[j](k\ell)}(\hat{B}) \rvert > \frac{\lambda_1 \omega_j n}{L^{1/2} K^{1/2}} \right] \\
        &\leq \PP\left[\max_{j\notin\{1,\dots,p_0\}} \max_{k\in\{1,\dots,K\}} \max_{\ell\in\{1,\dots,L\}} \lVert X_{[j](\cdot\ell)} \rVert_2 \lVert\Delta_k\rVert_2 > \frac{\lambda_1 \omega_j n}{2 L^{1/2} K^{1/2}} \right] \\
        &\leq \PP\left[\max_{j\notin\{1,\dots,p_0\}} \max_{k\in\{1,\dots,K\}} \max_{\ell\in\{1,\dots,L\}} \lVert \mathring{X}_{[j](\cdot\ell)} \rVert_2 \lVert\Delta_k\rVert_2 > \frac{\lambda_1 \omega_j n}{4 L^{1/2} K^{1/2}} \right] \\
        &+ \PP\left[\max_{j\notin\{1,\dots,p_0\}} \max_{k\in\{1,\dots,K\}} \max_{\ell\in\{1,\dots,L\}} \lVert {X}_{[j](\cdot\ell)} - \mathring{X}_{[j](\cdot\ell)} \rVert_2 \lVert\Delta_k\rVert_2 > \frac{\lambda_1 \omega_j n}{4 L^{1/2} K^{1/2}} \right] \\
        &\coloneqq P_1 + P_2\,.
    \end{split}
\end{equation*}

\noindent
As for $P_1$, notice that $\lVert \mathring{X}_{[j](\cdot\ell)} \rVert_2^2 = \sum_{i=1}^n \mathring{X}_{[j](i\ell)}^2 = \sum_{i=1}^n \langle\mathcal{X}_{ij}, \Xi_{j\ell} \rangle \leq \sum_{i=1}^n \cnorm{\mathcal{X}_{ij}}^2 \cnorm{\Xi_{j\ell}}^2 = O_p(n)$. Thus, $\max_{j\notin\{1,\dots,p_0\}} \max_{k\in\{1,\dots,K\}} \max_{\ell\in\{1,\dots,L\}} \lVert \mathring{X}_{[j](\cdot\ell)} \rVert_2 = O_p(n^{1/2})$. Therefore, by condition \textbf{E} and Corollary \ref{corollary:reduced_projection_rates}, we get:
\begin{equation*}
    \small
    \begin{split}
        \lVert \mathring{X}_{[j](\cdot\ell)} \rVert_2 \lVert\Delta_k\rVert_2 &= O_p(n^{1/2}(K^2 + Lp_0 + np_0L^{-2\gamma_2-\alpha_2})^{1/2}) \\
        &= o_p\left(\frac{\lambda_1 \omega_j n }{4L^{1/2}K^{1/2}}\right)\,,
    \end{split}
\end{equation*}
which implies $P_1 = o(1)$. \\

\noindent
As for $P_2$, by Lemma \ref{lemma:FPCrates}, we have $\lVert {X}_{[j](\cdot\ell)} - \mathring{X}_{[j](\cdot\ell)} \rVert_2^2 = O_p(L^2)$. Therefore, we get $\max_{j\notin\{1,\dots,p_0\}} \max_{k\in\{1,\dots,K\}} \max_{\ell\in\{1,\dots,L\}} \lVert {X}_{[j](\cdot\ell)} - \mathring{X}_{[j](\cdot\ell)} \rVert_2 = O_p(L)$, and, by condition \textbf{E}
\begin{equation*}
    \small 
    \begin{split}
        \lVert {X}_{[j](\cdot\ell)} - \mathring{X}_{[j](\cdot\ell)} \rVert_2 \lVert \Delta_k \rVert_2 &= O_p(L(K^2 + Lp_0 + np_0L^{-2\gamma_2-\alpha_2})^{1/2}) \\
        &= o_p\left(\frac{\lambda_1 \omega_j n}{4 K^{1/2} L^{1/2}}\right)\,,
    \end{split}
\end{equation*}
which implies $P_2 = o(1)$. 
 
}

%
%
\section{Additional figures and tables}
\label{sec:supp_add_sim}

\begin{algorithm}[h!]
\renewcommand{\thealgorithm}{E.1}
\caption{\textbf{FAStEN Implementation Specifics}}
\label{alg:implementation_detail}
\small
\begin{algorithmic}

\vspace{0.2cm}
\STATE \texttt{SELECT:} 
\begin{itemize}
    \vspace{-0.2cm}
    \item[-] Quantitative selection criterion (sc): \emph{gcv, cv}
    \vspace{-0.2cm}
    \item[-] Adaptive strategy: \emph{full, soft, null}
    \vspace{-0.2cm}
    \item[-] Basis system: \emph{unsupervised (X), supervised (Y)}
\end{itemize}

\STATE \texttt{SET:} $K=L=3$, $sc = \infty$.  
\vspace{0.2cm}

\STATE \texttt{DEFINE:} the optimal solution set $sol^{\text{opt}} = (B^{\text{opt}}, sc^{\text{opt}}, \lambda_1^{\text{opt}})$ 
\vspace{0.2cm}

\STATE \texttt{WHILE TRUE:}

\begin{itemize}
 
    \item[] $\ \cdot$ \texttt{find} matrix representation according to the selected basis system
    \vspace{-0.15cm}
    
    \item[] $\ \cdot$ \texttt{perform} $\lambda_1$ path-search with $\omega$'s equal to 1
    \vspace{-0.15cm}

    \item[] $\ \cdot$ \texttt{assign} to $sol^{\text{opt}}$ the solution that minimizes criterion computed with relaxed estimates

    \item[] \texttt{IF adaptive == soft:}
        \vspace{-0.3cm}
        \begin{itemize}
            \item[$\cdot$] \texttt{assign} to $sol^{\text{opt}}$ the solution of \texttt{FAStEN} with  $\omega_j=\text{sd}_B/\lVert \check{B}^R_{[j]}\rVert_2$ and $\lambda_1 = \lambda_1^{\text{opt}}$
        \end{itemize}

    \item[] \texttt{IF adaptive == full:}
        \vspace{-0.3cm}
        \begin{itemize}
            \item[$\cdot$] \texttt{perform} $\lambda_1$ path-search with  $ \omega_j=1/\lVert \check{B}^R_{[j]}\rVert_2$
            \vspace{-0.15cm}
            \item[$\cdot$] \texttt{assign} to $sol^{\text{opt}}$ the solution that minimizes criterion computed without relaxation
        \end{itemize}

    \item[] \texttt{IF $sc^{\text{opt}} < sc$:}  
        \vspace{-0.3cm}
        \begin{itemize}
            \item[$\cdot$] \texttt{set} $K=K+1$,  $L=L+1$ and $sc = sc^{\text{opt}}$ 
        \end{itemize}
    \vspace{-0.3cm}

     \item[] \texttt{ELSE:}  
        \vspace{-0.3cm}
        \begin{itemize}
            \item[$\cdot$]  \texttt{break} and \texttt{return} optimal solution of the previous $K$. 
        \end{itemize}
        \vspace{0.2cm}
\end{itemize}

\vspace*{-0.4cm}
\end{algorithmic}
\end{algorithm}

\begin{figure}[h]
    \centering
    \includegraphics[width=1\linewidth]{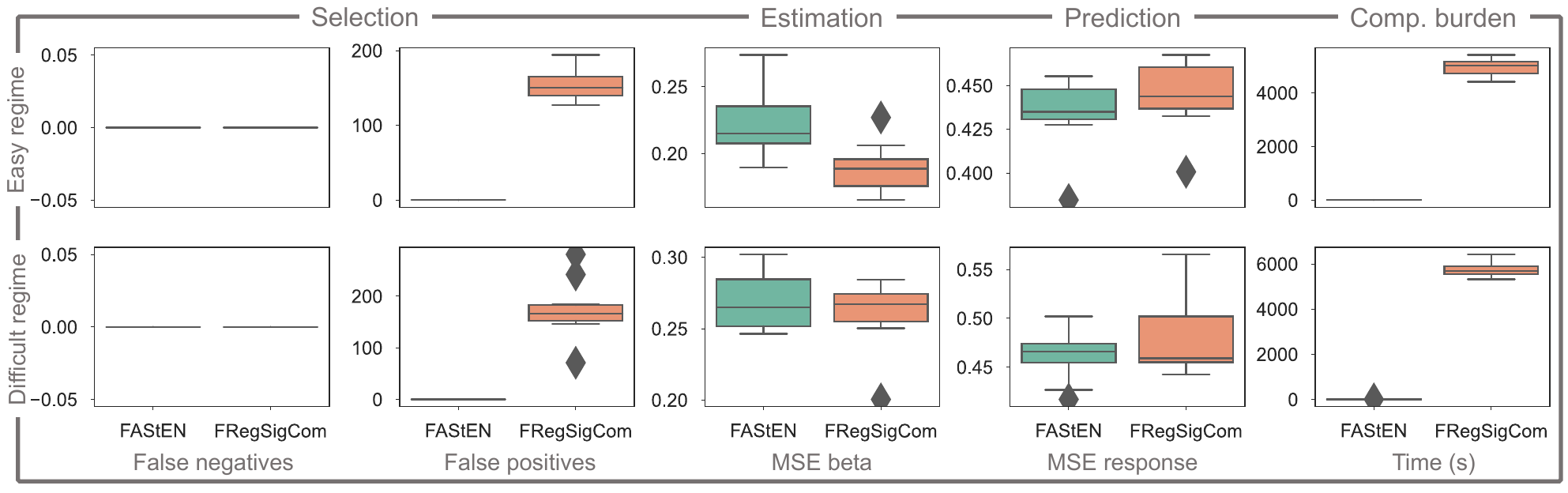}
    \vspace*{-0.6cm}
    \caption{
    \textbf{Benchmark vs. Competitor.}
    Performance comparison between \texttt{FAStEN} and \texttt{FRegSigCom}. Simulations were run 10 times with the following parameters: $n=300$, $p=500$, $p_0=10$, $snr=10$. Results are shown for ``easy'' (upper row) and ``difficult'' (lower row) coefficient scenarios.
    }

    \label{fig:supp_competitor}
\end{figure}

\begin{figure}[h]
    \centering
    \includegraphics[width=1\linewidth]{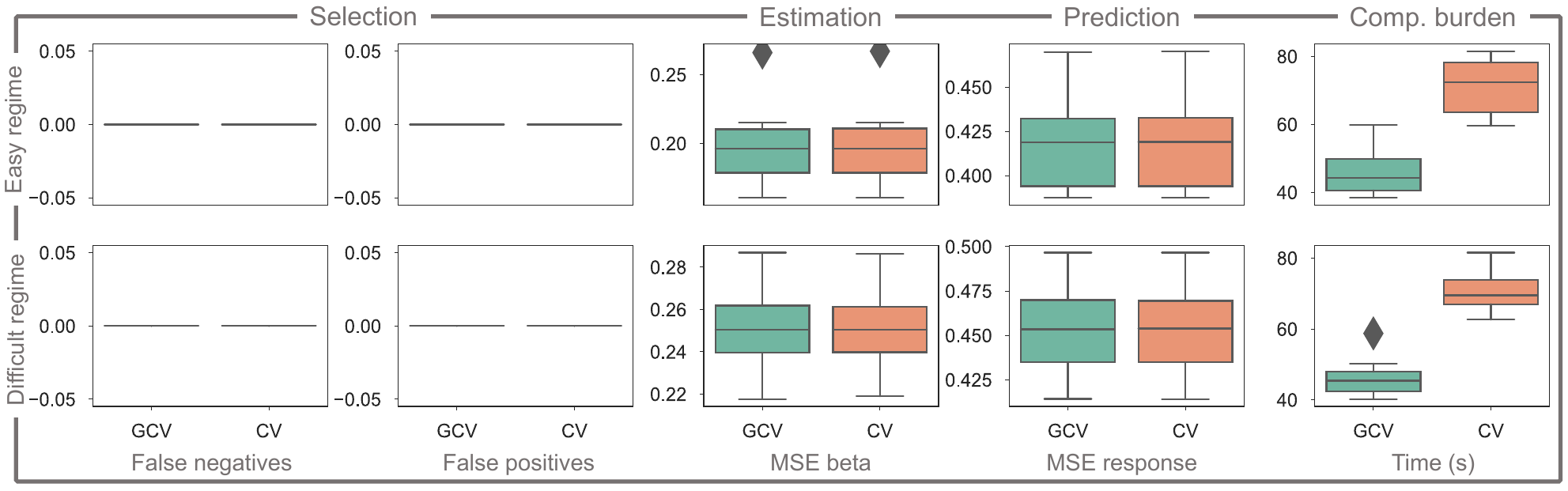}
    \vspace*{-0.6cm}
    \caption{
    \textbf{Quantitative Selection Criteria.}
    Performance comparison between \texttt{FAStEN} using \textit{gcv} and \textit{cv} as quantitative criteria. Simulations were run 10 times with the following parameters: $n=600$, $p=8000$, $p_0=10$, $snr=10$. Results are shown for ``easy'' (upper row) and ``difficult'' (lower row) coefficient scenarios.}
    \label{fig:supp_gcv_cv}
\end{figure}

\begin{figure}[h]
    \centering
    \includegraphics[width=1\linewidth]{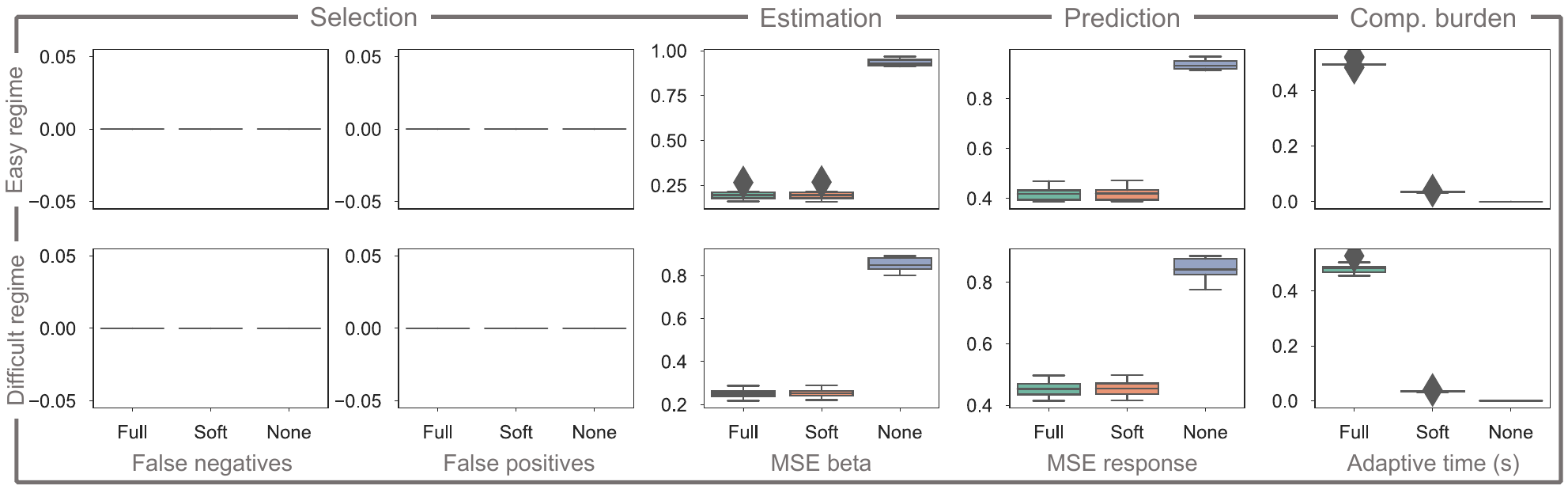}
    \vspace*{-0.6cm}
    \caption{
    \textbf{Adaptive Strategies.}
    Performance comparison between \emph{full adaptive}, \emph{soft adaptive}, and \emph{non-adaptive} \texttt{FAStEN} implementations. Simulations were run 10 times with the following parameters: $n=600$, $p=8000$, $p_0=10$, $snr=10$. Results are shown for ``easy'' (upper row) and ``difficult'' (lower row) coefficient scenarios. The computational burden illustrates only the marginal increase due to the adaptive implementation.}
    \label{fig:supp_adaptive_soft_none}
\end{figure}

\begin{figure}[h]
    \centering
    \includegraphics[width=1\linewidth]{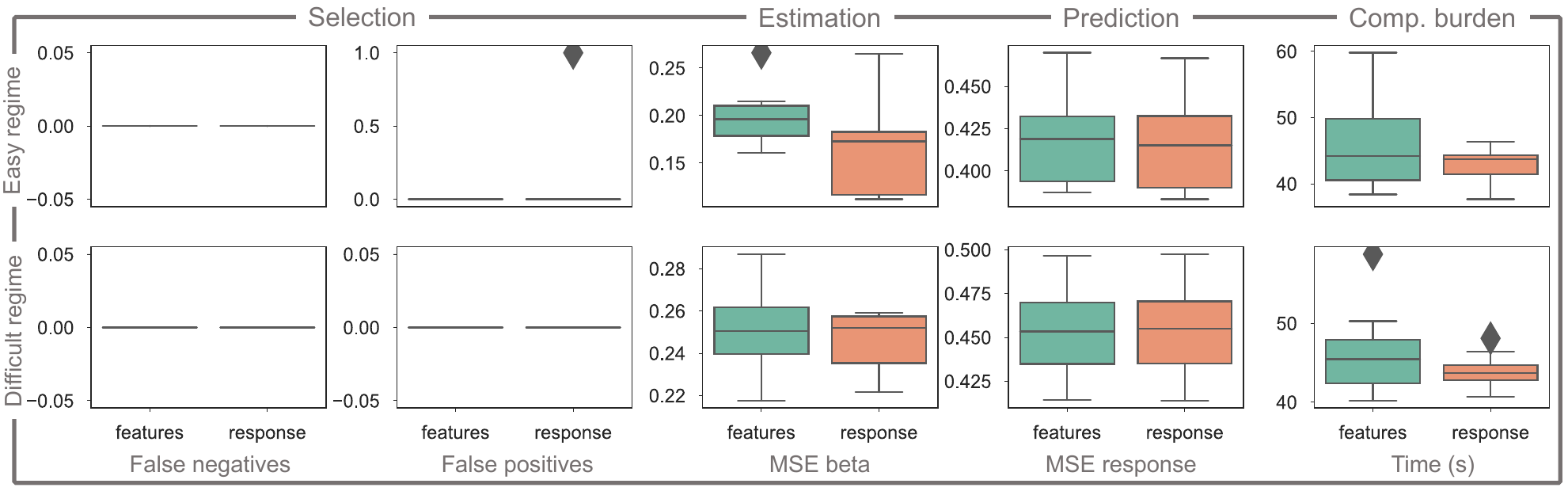}
    \vspace*{-0.6cm}
   \caption{
    \textbf{FPCs Reduction Strategies.}
    Performance comparison between \emph{supervised} and \emph{unsupervised} versions of \texttt{FAStEN}. Simulations were run 10 times with the following parameters: $n=600$, $p=8000$, $p_0=10$, $snr=10$. Results are shown for ``easy'' (upper row) and ``difficult'' (lower row) coefficient scenarios.}
    \label{fig:supp_X_Y}
\end{figure}

\begin{figure}[h]
    \centering
    \includegraphics[width=0.6\linewidth]{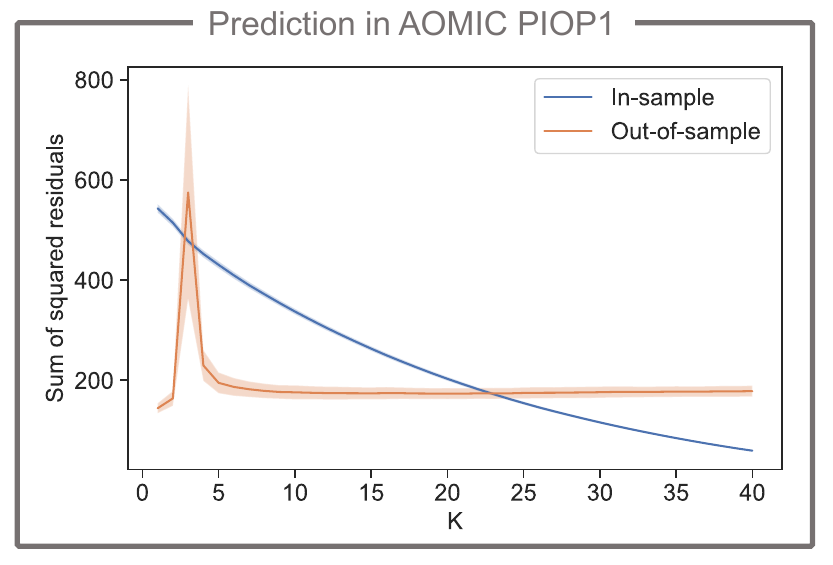}
   \caption{\textbf{Relationship between $K$ and prediction errors.} Observations from the AOMIC PIOP1 study are split into training (80\%) and test (20\%) sets by random sampling 30 times. Bold lines represent the average in-sample and out-of-sample prediction errors across replications. Shaded bands indicate $\pm1$ standard deviation intervals across the 30 replications. As $K$ increases, the in-sample prediction error decreases, while the out-of-sample prediction error stabilizes at a constant level.}

    \label{fig:supp_app_r2}
\end{figure}


\end{document}